# Prototype-guided transfer of sparse literature knowledge for electrolyte additive discovery

Weixiang HONG[1,5], Hongting DU[1,5], Jiayue TANG[1], Ruifeng TAN[1], Yangjian QUAN[2], Jia LI[3,*], Jiaqiang HUANG[1,4,*]

[1] Sustainable Energy and Environment Thrust and Guangzhou Municipal Key Laboratory of Materials Informatics, The Hong Kong University of Science and Technology (Guangzhou), Nansha, Guangzhou, 511400, Guangdong, P.R. China
[2] Department of Chemistry, The Hong Kong University of Science and Technology, Clear Water Bay, Kowloon, Hong Kong SAR, P.R. China
[3] Data Science and Analytics Thrust and Guangzhou Municipal Key Laboratory of Materials Informatics, The Hong Kong University of Science and Technology (Guangzhou), Nansha, Guangzhou, 511400, Guangdong, P.R. China
[4] Academy of Interdisciplinary Studies, The Hong Kong University of Science and Technology, Clear Water Bay, Kowloon, Hong Kong SAR, P.R. China
[5] These authors contributed equally: Weixiang HONG, Hongting DU

* Corresponding authors:
J. Huang: seejhuang@hkust-gz.edu.cn
J. Li: jialee@hkust-gz.edu.cn

## Abstract

Electrolyte additive discovery remains challenging because experimentally validated molecules are sparse, whereas accessible chemical spaces are vast and largely unlabeled. This challenge is amplified in lithium-ion batteries, where additive performance arises from coupled interfacial reactions rather than a single molecular property. Here, we develop a prototype-guided molecular intelligence, ProtoMI, a literature-driven framework that learns transferable structural priors from reported electrolyte additives and uses them to prioritize candidates in unlabeled chemical space. For boron-containing additives, ProtoMI combines 126 literature-reported molecules with 179,977 unlabeled candidates. Graph contrastive learning identifies seven chemically interpretable prototypes from the reported additives, and prototype-guided semi-supervised contrastive learning adapts these prototypes to the candidate space under source-target distribution mismatch. In retrospective temporal validation, ProtoMI achieves enrichment factors of 9.2-45.6 while screening less than 2% of the candidate space. A subsequent translation step identifies four commercially accessible candidates. One representative candidate, 4,4,5,5-Tetramethyl-2-[10-(1-naphthyl)anthracen-9-yl]-1,3,2-dioxaborolane (TNDB), improves high-temperature $LiFePO_4$||graphite cycling at 55 °C by 34.93% relative to the baseline electrolyte. An arsenal of characterizations and operando optical fiber Fourier transform infrared spectroscopy suggest that TNDB forms B-containing, F/P/O-modified inorganic interphases, suppresses solvent decomposition and reduces Fe deposition on graphite. This case study shows how sparse literature knowledge can guide experimentally efficient molecular discovery in data-scarce battery-additive spaces.

## Introduction

Lithium-ion batteries (LIBs) underpin electrified transport, grid storage and portable electronics, and their deployment is expected to expand substantially as electrification accelerates[1,2]. As LIBs are pushed toward longer lifetimes, higher power operation and harsher operating environments, battery degradation is increasingly governed by the stability of electrochemical interfaces, where electrolyte decomposition, interphase evolution and transition-metal crossover can trigger coupled degradation processes[3-5]. Electrolyte additives provide a practical and manufacturable route to regulate such interfaces. Even at low concentrations, additives can alter solvation structures, undergo preferential interfacial reactions, scavenge reactive impurities and promote stabilizing solid-electrolyte interphase (SEI) and cathode-electrolyte interphase (CEI)[6]. However, additive discovery remains largely empirical because additive efficacy emerges from the coupled effects of molecular stability, salt-solvent coordination, ion transport, decomposition pathways and interphase growth, none of which can be reliably captured by a single molecular descriptor[3-6].

Recent advances in artificial intelligence (AI), molecular representation learning and data-driven optimization have accelerated the exploration of high-dimensional chemical spaces[7,8]. In battery-electrolyte research, data-driven strategies have been used[9] to predict molecular and electrolyte properties[10], screen electrolyte components[11] with theoretical calculations[12-16] or curated datasets[17,18], and guide electrolyte optimization workflows[18-20]. Active-learning approaches have further reduced experimental burden by iteratively updating models with new measurements[20], while physics-informed and solvation-aware models have introduced thermodynamic, transport and molecular-interaction constraints into electrolyte design[21,22]. These studies have demonstrated the value of data-driven electrolyte discovery. Nevertheless, most existing approaches are formulated around supervised prediction or iterative optimization once target labels, simulation outputs or experimental feedback are available. This creates a mismatch for electrolyte additive discovery, where experimentally validated molecules are sparse, literature reports are heterogeneous and most accessible candidate molecules have no measured battery-relevant labels.

Under such severe label scarcity, the central challenge shifts from predicting the performance of individual molecules to prioritizing chemically promising regions of an unlabeled molecular space under limited experimental budgets. Reported electrolyte additives are not merely isolated successful examples. Instead, they may encode recurring structural organization and shared interfacial regulation strategies[6], including anion reception[23], reactive-species scavenging[24], solvation regulation[25] and interphase construction[24,26]. If these recurring patterns can be learned as prototype-level priors rather than as molecule-level labels, they may guide exploration beyond the local neighborhoods of known additives. Representation learning provides a route to extract structured molecular embeddings from sparse data[27], while prototype-based alignment[28,29] and semi-supervised learning[30] offer mechanisms to transfer such organization into unlabeled domains. Together, these advances suggest an alternative discovery paradigm: instead of directly predicting additive performance from scarce labels, sparse literature-derived success patterns can be converted into transferable molecular prototypes that guide efficient candidate selection (Fig. 1a,b).

Here we develop prototype-guided molecular intelligence (ProtoMI), a literature-driven framework for molecular discovery under sparse positive knowledge and large unlabeled candidate spaces. ProtoMI treats reported additives as a positive molecular knowledge source rather than as a conventional supervised training set. The framework proceeds through three stages (Fig. 1c). First, literature-reported additive molecules are extracted and curated to construct a positive molecular dataset, and the unsupervised graph contrastive learning is used to identify prototype-level chemical organization within the reported additives. Second, prototype-guided semi-supervised contrastive learning adapts these prototypes to a large unlabeled molecular space under source-target distribution mismatch. Third, a subsequent knowledge-guided translation step incorporates chemical feasibility and experimental accessibility constraints to convert model recommendations into testable candidates.

As a case study, we focus on boron-containing electrolyte additives. Boron-containing additives provide a suitable testbed because their Lewis-acidic and borate/boronate chemistries are well documented in electrolyte interphase regulation, yet remain sparsely explored across the broader boron-containing chemical space[23-26]. This combination of established chemical relevance and large unexplored molecular diversity creates a stringent test for ProtoMI: whether sparse literature-reported successes can be transformed into transferable prototype priors for efficient candidate selection. In this work, ProtoMI integrates 126 literature-reported boron-containing additives with 179,977 unlabeled boron-containing candidates from PubChem[31]. Retrospective temporal validation shows that the framework recovers future-reported additives with enrichment factors of 9.2-45.6 while screening less than 2% of the candidate space. A knowledge-guided translation step further prioritizes four commercially accessible candidates for high-temperature $LiFePO_4$||graphite cells, a system in which elevated-temperature cycling stability remains an important practical challenge[32,33]. One representative candidate, 4,4,5,5-tetramethyl-2-[10-(1-naphthyl)anthracen-9-yl]-1,3,2-dioxaborolane (TNDB), improves capacity retention by 34.93% relative to the baseline electrolyte during cycling at 55 °C. Post-mortem and operando characterizations indicate that TNDB participates in forming B-containing, F/P/O-modified interphases that suppress solvent decomposition and reduce Fe deposition on graphite. These results demonstrate that sparse literature-derived molecular successes can be transformed into transferable prototype priors for efficient electrolyte additive discovery under severe data scarcity.

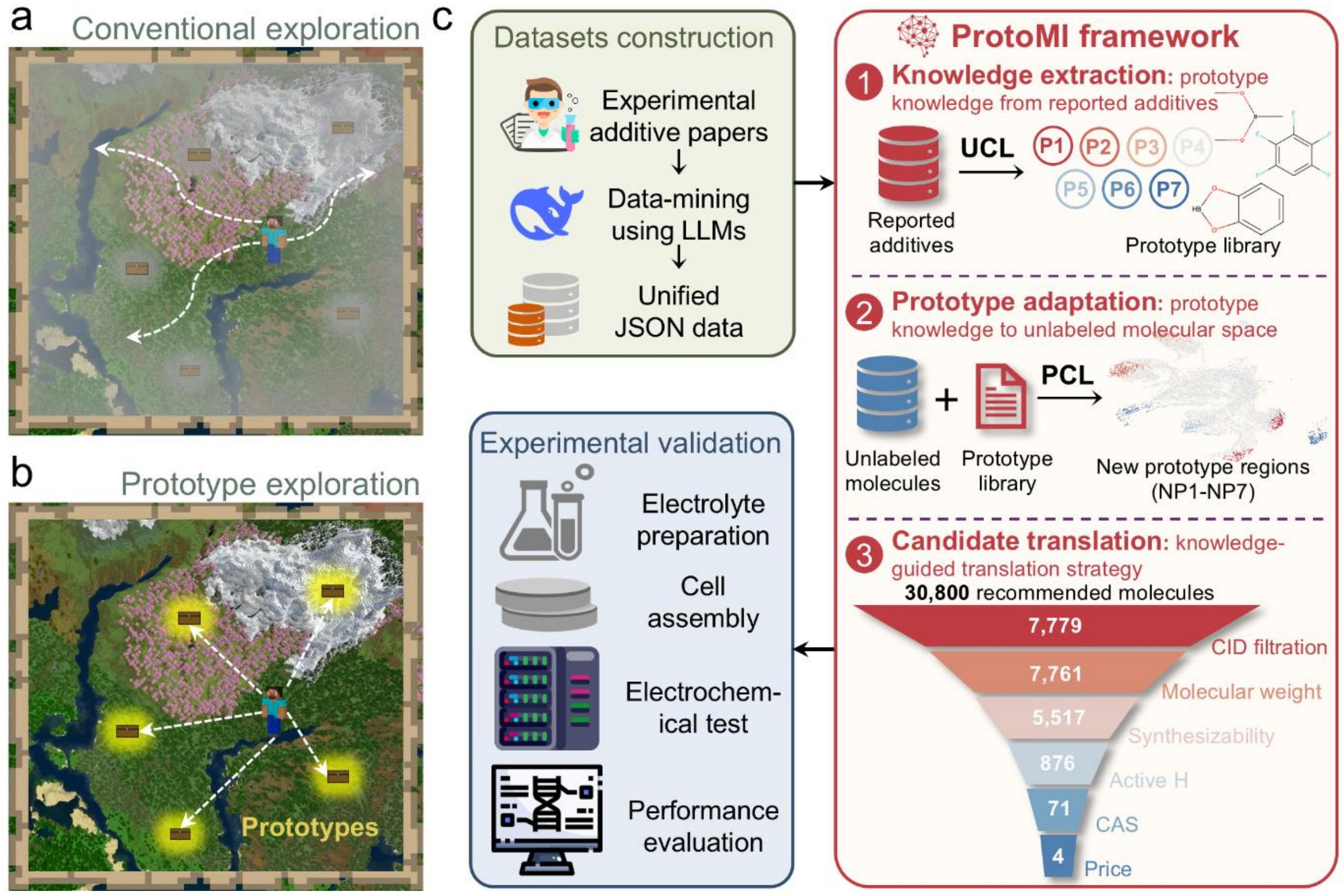

**Fig. 1 | Overview of the ProtoMI framework for prototype-guided discovery of boron-containing electrolyte additives.** **a**,**b**, Conceptual comparison between traditional molecular exploration and prototype-guided exploration. Conventional exploration (**a**) searches the molecular space largely through local trial-and-error optimization, whereas prototype-guided exploration (**b**) first identifies representative prototype regions from reported additives and subsequently explores structurally distinct regions of the molecular landscape through prototype-level knowledge transfer. **c,** Overview of the ProtoMI molecular discovery workflow. Experimentally reported boron-containing additives are first collected from the literature using a large-language-model-assisted data-mining pipeline to construct a curated molecular dataset. An unsupervised graph contrastive learning model then extracts prototype-level chemical knowledge from reported additives. These prototypes are transferred to a large unlabeled molecular space by prototype-guided semi-supervised contrastive learning, yielding adapted prototype regions and recommended candidates. A knowledge-guided translation step further incorporates chemical feasibility and practical accessibility constraints to generate experimentally actionable candidates for validation.

## Results

### Dataset construction reveals source-target mismatch in boron-additive chemical space

We first formulated boron-containing electrolyte additive discovery as a sparse-positive molecular exploration problem. To define the target search space, we collected 179,977 boron-containing candidate molecules from PubChem[31]. In parallel, we constructed a positive dataset of 126 experimentally reported boron-containing electrolyte additives extracted from Web of Science articles using a large language model (LLM)-assisted heuristic pipeline (Fig. 2a and Methods M1). This design reflects a realistic discovery setting: successful additives are sparsely reported in the literature, whereas the accessible boron-containing chemical space is orders of magnitude larger and remains largely unlabeled.

To evaluate whether unstructured literature could provide reliable positive molecular knowledge, we manually verified the extracted information across multiple battery-relevant fields. The extraction pipeline achieved high accuracy for chemically important categories, including 94% accuracy for additive identification, while requiring approximately 11 s per article and less than 60 CNY in total processing cost (Fig. 2b). These results indicate that LLM-assisted literature extraction can provide an efficient route to constructing a curated positive molecular set for downstream prototype learning. Overall dataset statistics are summarized in Supplementary Table S1.

The resulting datasets reveal a pronounced source-target distribution mismatch between literature-reported additives and the unlabeled boron-containing candidate space. Uniform manifold approximation and projection (UMAP)[34] based on Morgan fingerprint[35] representations shows that reported additives occupy restricted regions within the broader unlabeled molecular landscape rather than being uniformly distributed across it (Fig. 2c). Hierarchical clustering[36] dendrogram further suggests that the reported additives contain recurring molecular patterns rather than isolated successful examples (Supplementary Fig. S1). Consistently, the molecular-weight–synthetic accessibility (SA)[37] map shows that positive molecules are concentrated mainly in low-to-moderate molecular-weight regions with moderate SA scores, whereas the unlabeled candidates span a much broader and more heterogeneous chemical space (Fig. 2d).

Graph-level descriptors provide additional evidence for population-level differences between the positive and unlabeled sets. Reported additives and unlabeled candidates differ in molecular size and connectivity, as reflected by shifts in atom number, bond number, average degree, graph density and degree entropy (Fig. 2e and Methods M2). Node-feature composition analysis further reveals feature-level differences between the two populations, suggesting that reported additives are enriched in specific local chemical environments. Representative shifted node features include scaled van der Waals radius (N67), scaled covalent radius (N68), aromaticity indicator (N65), $sp^2$ hybridization (N59) and carbon atom identity (N00) (Fig. 2f and Supplementary Table S2).

Together, these analyses define both the challenge and the opportunity of data-scarce additive discovery. Experimentally reported additives are too sparse and distributionally biased to support exhaustive or purely supervised screening, yet they exhibit non-random structural organization that may encode transferable chemical priors. We therefore next asked whether this organization could be learned as prototype-level knowledge and used to guide exploration of the broader unlabeled candidate space.

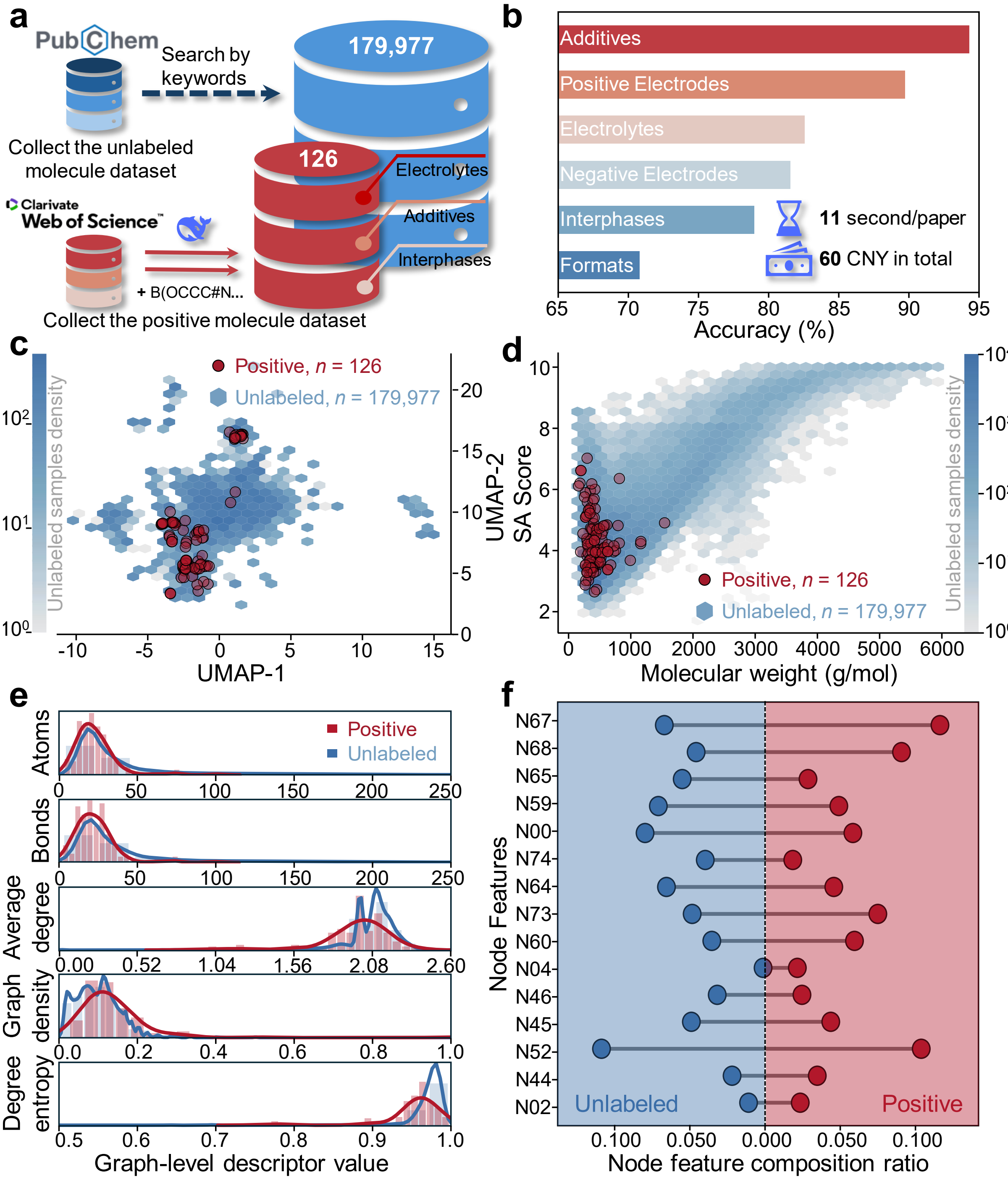

**Fig. 2 | Dataset construction and source-target mismatch in boron-additive chemical space. a,** Workflow for constructing the molecular datasets. The unlabeled search space consists of 179,977 boron-containing candidate molecules collected from PubChem, whereas the positive dataset contains 126 literature-reported boron-containing electrolyte additives extracted from Web of Science articles using an LLM-assisted heuristic pipeline. **b,** Manual validation of the extraction pipeline across battery-relevant fields, showing reliable extraction of chemically important information. **c,** UMAP projection of Morgan fingerprint representations. Positive molecules are localized within restricted regions of the

unlabeled molecular landscape, while background hexagons represent the logarithmic density of unlabeled candidates. **d,** Molecular-weight and SA distribution of positive and unlabeled molecules. Positive additives are enriched in low-to-moderate molecular-weight regions with moderate SA scores, whereas unlabeled candidates cover a broader chemical space. Molecular weights are shown within 0-6,000 g $mol^{-1}$, as higher values are extremely rare (<0.1% of the dataset). **e,** Graph-level descriptor distributions reveal differences in molecular size (Atoms and Bonds), connectivity (Average degree) and topology (Graph density and Degree entropy) between the two molecular populations. **f,** Mirrored dumbbell plot showing normalized composition ratios of the most shifted node features, highlighting feature-level differences between positive and unlabeled molecules.

### Reported additives encode transferable molecular prototypes

The clustered organization of the positive dataset suggests that successful boron-containing additives may share recurring structural principles rather than represent isolated molecular examples. To extract this organization, we applied an unsupervised graph contrastive learning (UCL) model[38] to learn prototype-level representations from the 126 reported additives. Reported molecules were converted into molecular graphs and expanded into augmented graph views using graph augmentation operations[39-43], generating 756 augmented graph samples from the original positive set (Fig. 3a and Methods M3). Paired graph views were then encoded by a graph neural-network encoder and optimized with the InfoNCE loss to learn augmentation-invariant molecular representations. The encoder consisted of repeated GINEConv, GraphNorm, ReLU and dropout layers for hierarchical graph representation extraction (Fig. 3b).

We first compared different molecular representation methods using clustering quality and computational cost as evaluation metrics (Fig. 3d). Conventional fingerprint-based[44] and SMILES-based[45] representations produced relatively low silhouette scores[46] of 0.258-0.337, indicating limited separation of reported additive groups in representation space. Graph-based encoders, including graph convolutional networks (GCN)[47], graph attention networks (GAT)[48] and graph isomorphism networks with edge features (GINE)[49], substantially improved cluster separability, with silhouette scores of 0.751-0.832. Augmentation-enhanced graph representations further improved embedding organization relative to the corresponding non-augmented encoders. Among all tested methods, GINE+ achieved the highest silhouette scores using both K-Means[50] clustering and hierarchical clustering, with values of 0.853 and 0.881, respectively. These results indicate that augmentation-invariant graph representations capture hidden structural organization within the reported additive set.

Hierarchical clustering of the learned GINE+ representation space identified seven molecular prototypes (Fig. 3c and Supplementary Figs. S2,S3). These prototypes displayed non-uniform motif distributions (Fig. 3e), indicating that they are organized by distinct combinations of recurring boron-related molecular fragments. For example, P3 was enriched in oxalate-containing and tetra-alkoxy frameworks, whereas P2 preferentially contained fluorinated aromatic fragments. P4 was characterized by catechol-derived structures, while P6 and P7 were dominated by fluorinated alkoxy and tri-alkoxy frameworks, respectively. These motif-level differences suggest that the learned prototypes capture chemically meaningful structural preferences rather than arbitrary clusters (Supplementary Tables S3,S4).

To further interpret the prototype space, we annotated the learned clusters using literature-supported functional categories and representative molecules (Fig. 3c, Supplementary Fig. S4 and Table S5). P1 showed relatively balanced contributions from interphase formation, solvation regulation and polymer strengthening, represented by the boron-containing heterocycle DiOB-Py, which has been reported to act as a mild anion receptor and facilitate interfacial charge transfer[25]. P2 combined Lewis acidity, solvation regulation and interphase formation, represented by tris(pentafluorophenyl)borane, whose fluorinated aromatic framework supports anion reception and interfacial stabilization[23]. P3 displayed a strong interface-engineering signature, represented by Lithium difluoro(oxalato)borate (LiDFOB), which is known to form F- and B-rich interphases that suppress parasitic reactions and stabilize electrode surfaces[26]. P4-P6 showed complementary stabilization-oriented profiles, represented by lithium bis[1,2-benzenediolato(2-)-O,O']borate (LBBB), 3-Pyridinecarbonitrile BF (PCN) and tris(2,2,2-trifluoroethyl)borate (TTFEB), which illustrate weakly coordinating borate chemistry, Lewis acid-base interactions and fluorinated anion-receptor behavior[51-53]. P7 was associated with solvation regulation and polymer strengthening, represented by tributyl borate, which promotes salt dissociation and boron-rich interphase formation in polymerized electrolyte systems[54].

These results show that the positive additive set can be converted into a prototype library that captures both recurring structural motifs and literature-supported functional tendencies. Importantly, the prototypes are learned without direct performance labels, making them suitable as transferable priors for candidate exploration under severe label scarcity. We therefore used this prototype library as the source knowledge for adaptation to the unlabeled boron-containing molecular space.

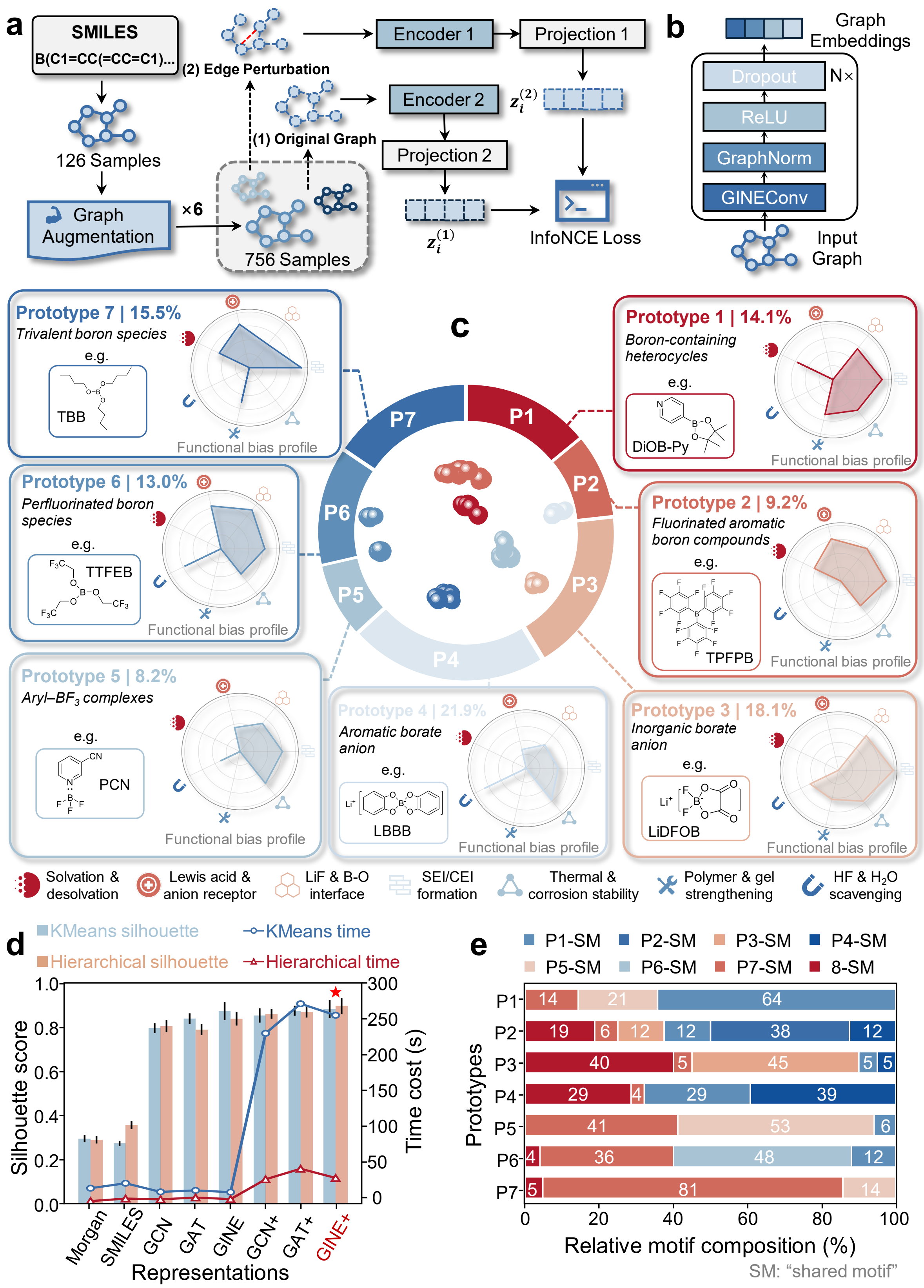

**Fig. 3 | Interpretable molecular prototypes emerge from reported boron-containing electrolyte additives. a,** Workflow of the unsupervised graph contrastive learning model. Reported additives are converted from SMILES strings into molecular graphs and expanded into multiple augmented graph views using five graph augmentation strategies. Paired graph views are encoded by graph neural-network encoders and projection heads, followed by InfoNCE optimization to learn augmentation-invariant molecular representations. **b,** Structure of the molecular graph encoder. The encoder consists

of repeated GINEConv, GraphNorm, ReLU and dropout layers, producing graph-level embeddings for clustering and prototype discovery. In this work, the repeated $N$ equals 3. **c,** Functional interpretation of the learned prototype space. Seven chemically distinct prototypes are identified from reported additives. The surrounding panels show characteristic additive families, representative molecules and functional bias profiles across seven electrolyte-regulation mechanisms. **d,** Comparison of representation methods in terms of clustering quality (K-Means and hierarchical clustering) and computational cost. Silhouette scores quantify the separability of molecular clusters. Methods labeled with "+" indicate augmentation-enhanced representations. **e,** Relative composition of representative boron-related motifs across the seven learned prototypes. The motif-level distribution shows that different prototypes are associated with distinct substructural preferences, suggesting that the learned representation space captures chemically meaningful organization. See the shared-motif summary table in Supplementary Table S4.

## Prototype-guided adaptation transfers additive knowledge to unlabeled chemical space

Although the prototype library extracted from reported additives captures transferable chemical organization, it does not by itself determine how limited experimental resources should be allocated within the much larger unlabeled candidate space. To address this challenge, we developed a prototype-guided semi-supervised contrastive learning (PCL) model, as the adaptation stage of ProtoMI (Fig. 4a and Methods M4). The PCL module combines exponential moving average (EMA)[55] updating of prototype centroids, top-k prototype-relevant sample assignment and prototype decorrelation regularization. Together, these components are designed to transfer prototype-level knowledge from the positive additive domain to the unlabeled candidate domain while maintaining both recommendation quality and prototype diversity.

During training, the prototype centroids exhibited a characteristic three-stage trajectory, consisting of initial exploration, progressive refinement and eventual stabilization (Fig. 4b). The gradual reduction in prototype drift indicates that the model converges toward a stable prototype organization in the adapted molecular representation space. This behavior supports the use of adapted prototypes as anchors for candidate prioritization.

We next examined how prototype knowledge was transferred from the reported additive space to the unlabeled candidate space. The correspondence between original prototypes P1-P7 and adapted, new prototypes NP1-NP7 shows that adaptation is not a simple one-to-one copying process (Fig. 4c). Instead, each adapted prototype integrates contributions from multiple original prototype regions, indicating that the PCL model reorganizes literature-derived knowledge according to the structure of the target candidate space while preserving prototype-level continuity. Consistently, the cosine-similarity map between initial and adapted prototypes shows strong diagonal correspondence together with structured off-diagonal relationships (Supplementary Fig. S5). Motif analysis further shows that characteristic boron-containing fragments are retained and redistributed across adapted prototypes, supporting chemically meaningful transfer rather than arbitrary re-clustering (Supplementary Fig. S6).

We then benchmarked the complete ProtoMI framework against multiple recommendation strategies (Table 1 and Fig. 4d). ProtoMI achieved high prototype similarity, assignment entropy and cluster separability, with values of 0.789 ± 0.030, 0.723 and 0.976, respectively. This indicates that recommended molecules remain both structurally coherent (Supplementary Fig. S7) and broadly distributed across multiple prototype regions (Supplementary Fig. S8). The corresponding entropy-weighted prototype similarity (Methods M5) reached 0.57 (or 57%), exceeding the competing recommendation pipelines. These results indicate that ProtoMI selects molecules that remain close to learned prototype regions while preserving coverage across multiple chemically distinct directions, thereby avoiding collapse into a few dominant structural families.

Ablation analysis further clarifies the contribution of each PCL component (Fig. 4e). Removing top-k assignment reduces the model's ability to select high-confidence molecules most compatible with each prototype. Removing prototype decorrelation decreases inter-prototype separation and promotes redundancy among adapted prototype directions. Removing EMA updating destabilizes cross-domain adaptation by allowing abrupt prototype shifts during training. The performance degradation observed after removing these components indicates that effective prototype adaptation requires coordinated control of sample confidence, prototype diversity and transfer stability (Supplementary Fig. S9). The conclusion is further supported by additional statistics in Supplementary Table S6.

Finally, we evaluated whether prototype adaptation improves discovery efficiency using retrospective temporal validation. Across four temporal splits using 2017, 2019, 2021 and 2023 as cutoff years, ProtoMI recovered future-reported additives with recall rates of 8.8-20.0% while screening only 0.19-1.27% of the candidate space, corresponding to enrichment factors of 9.2-45.6 relative to random screening (Fig. 4f and Supplementary Table S7). Pareto-front analysis further shows that ProtoMI concentrates discovery efforts into a small fraction of the search space while preserving substantial recall, corresponding to effective search-space reductions of 79-517 folds (Fig. 4g). These results demonstrate that sparse literature-derived prototype knowledge can be transferred into unlabeled chemical space to improve candidate prioritization under limited experimental budgets.

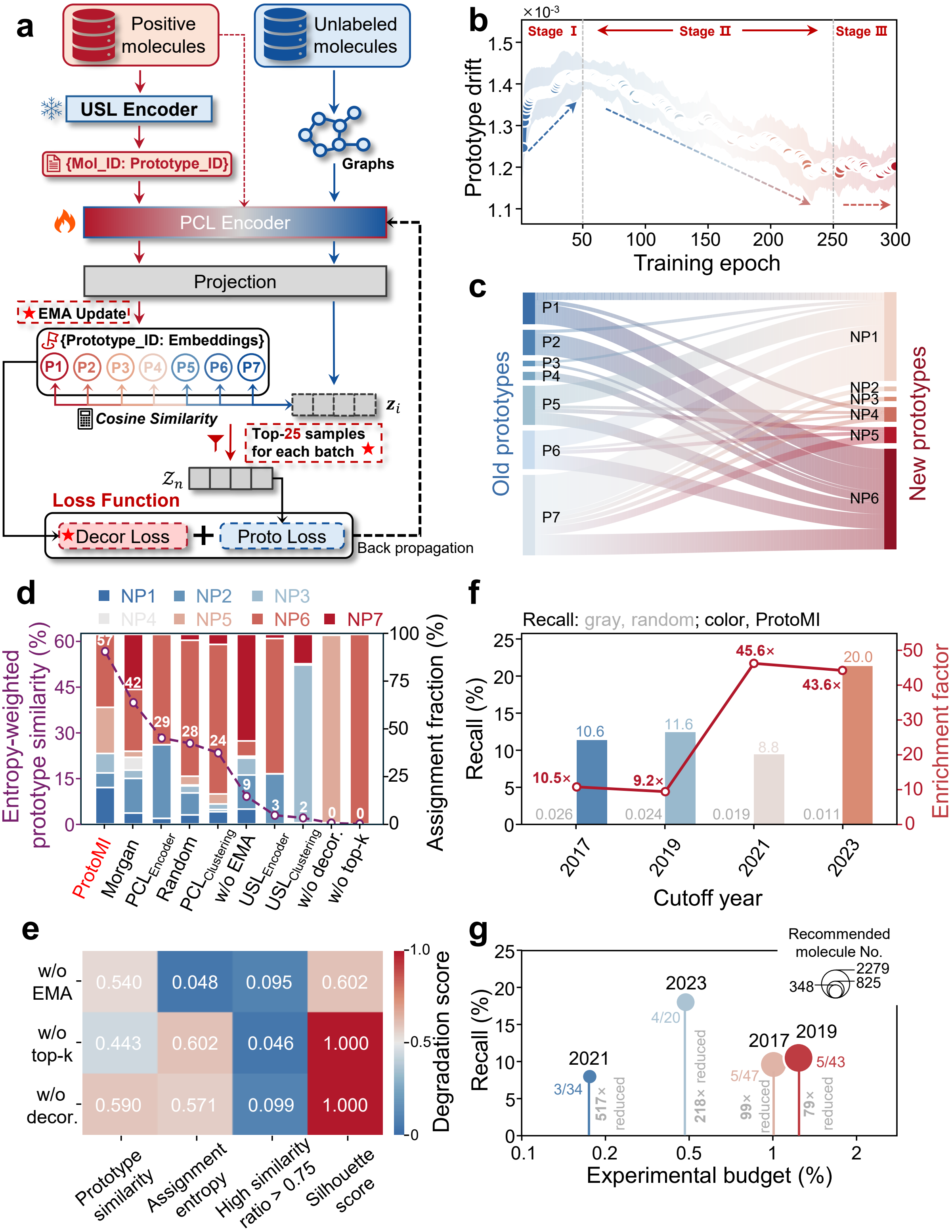

**Fig. 4 | Prototype adaptation transfers additive knowledge to unlabeled boron-containing chemical space. a,** Workflow of the PCL module within ProtoMI. Molecular prototypes extracted from reported boron-containing electrolyte additives are used to guide representation learning and candidate recommendation in the unlabeled molecular space. **b,** Mean evolution of prototype drift during training

across 10 trials. Prototype centroids undergo an initial exploration stage (Stage I), followed by progressive refinement (Stage II) and eventual stabilization (Stage III), revealing the emergence of a converged prototype organization in the learned molecular representation space. **c,** Knowledge retention between original prototypes and adapted prototypes. The Sankey diagram shows the redistribution of structural motifs from reported-additive prototypes to newly formed prototypes in unlabeled molecular space. **d,** Entropy-weighted prototype similarity and prototype distribution of molecules selected by different recommendation pipelines. ProtoMI maintains high prototype consistency while preserving coverage across multiple prototypes. **e,** Ablation analysis of key components in the PCL module. Removing EMA updating, prototype decorrelation or top-k prototype selection degrades recommendation quality, highlighting their roles in stable prototype formation and effective knowledge transfer. **f,** Temporal validation using historical splits. Recall rates are compared with random screening, and the red line indicates the enrichment factor achieved under limited experimental budgets. **g,** Pareto front of discovery efficiency. Bubble size represents the number of recommended molecules, illustrating the trade-off between recall ability and experimental budget across different historical splits. "*x*/*y*" in the figure indicates the correct prediction of *x* molecules among the literature reported *y* molecules.

**Table 1 | Recommendation performance of ProtoMI and baseline strategies**

| Category | Method | Similarity | Entropy | Silhouette |
|---|---|---|---|---|
| Random | Random selection | 0.521 ± 0.147 | 0.529 | 0.596 |
| Fingerprint | Morgan fingerprints | 0.497 ± 0.177 | **0.836** | 0.408 |
| Encoder | UCL encoder only | 0.085 ± 0.010 | 0.349 | -0.185 |
| | PCL encoder only | 0.728 ± 0.045 | 0.404 | 0.968 |
| Clustering | UCL encoder clustering | 0.079 ± 0.021 | 0.260 | 0.786 |
| | PCL encoder clustering | 0.571 ± 0.110 | 0.426 | 0.916 |
| Prototype-guided | ProtoMI | **0.789±0.030** | 0.723 | **0.976** |
| Ablation | w/o EMA | 0.141 ± 0.036 | 0.669 | 0.253 |
| | w/o decorrelation | 0.080 ± 0.014 | 0.037 | -0.224 |
| | w/o top-k | 0.257 ± 0.000 | 0.000 | N/A |

Bold and gray-shaded values indicate the best and second-best performance, respectively.

### Knowledge-guided translation converts recommendations into testable candidates

Prototype adaptation reduced the original search space from 179,977 boron-containing candidates to 30,800 ProtoMI-recommended molecules. However, many recommended molecules remained chemically impractical or experimentally inaccessible. To bridge the gap between computational prioritization and laboratory validation, we developed a knowledge-guided translation strategy that incorporates empirical additive-design constraints and practical accessibility requirements (Fig. 5a and Methods M6).

The translation workflow sequentially applied filters based on PubChem CID validity, molecular-weight range, synthetic accessibility, active-hydrogen exclusion, Chemical Abstracts Service (CAS) availability and practical cost. This process reduced the recommendation pool from 30,799 molecules to 68 supplier-accessible candidates, which were further prioritized according to purchasability, cost and prototype traceability (Supplementary Fig. S10 and Supplementary Table S8). Four low-cost commercially available molecules were selected for experimental consideration (Supplementary Table S9). This workflow separates the model-driven recommendation stage from the practical translation stage, ensuring that experimentally selected molecules are both chemically relevant and accessible under realistic resource constraints.

To assess whether prototype-level organization was preserved during translation, we projected the retained molecules onto the adapted representation landscape (Fig. 5b). Post-screened molecules remained distributed across multiple prototype regions, indicating that practical filtering did not collapse the recommendations into a single structural family. No commercially accessible molecules were retained from NP4 and NP7 after final filtering, although candidate molecules from these regions existed before post-screening. This suggests that their absence among final candidates reflects practical accessibility constraints rather than a lack of prototype coverage. Representative unselected molecules, UM#1-UM#3, are shown to illustrate structurally plausible molecules that were present in the unlabeled pool but not retained as final experimental candidates.

Together, these results show that knowledge-guided translation serves as a practical bridge between prototype-guided molecular exploration and experimental implementation. By incorporating feasibility and accessibility constraints after model recommendation, ProtoMI converts a large prioritized molecular pool into a small set of experimentally actionable candidates.

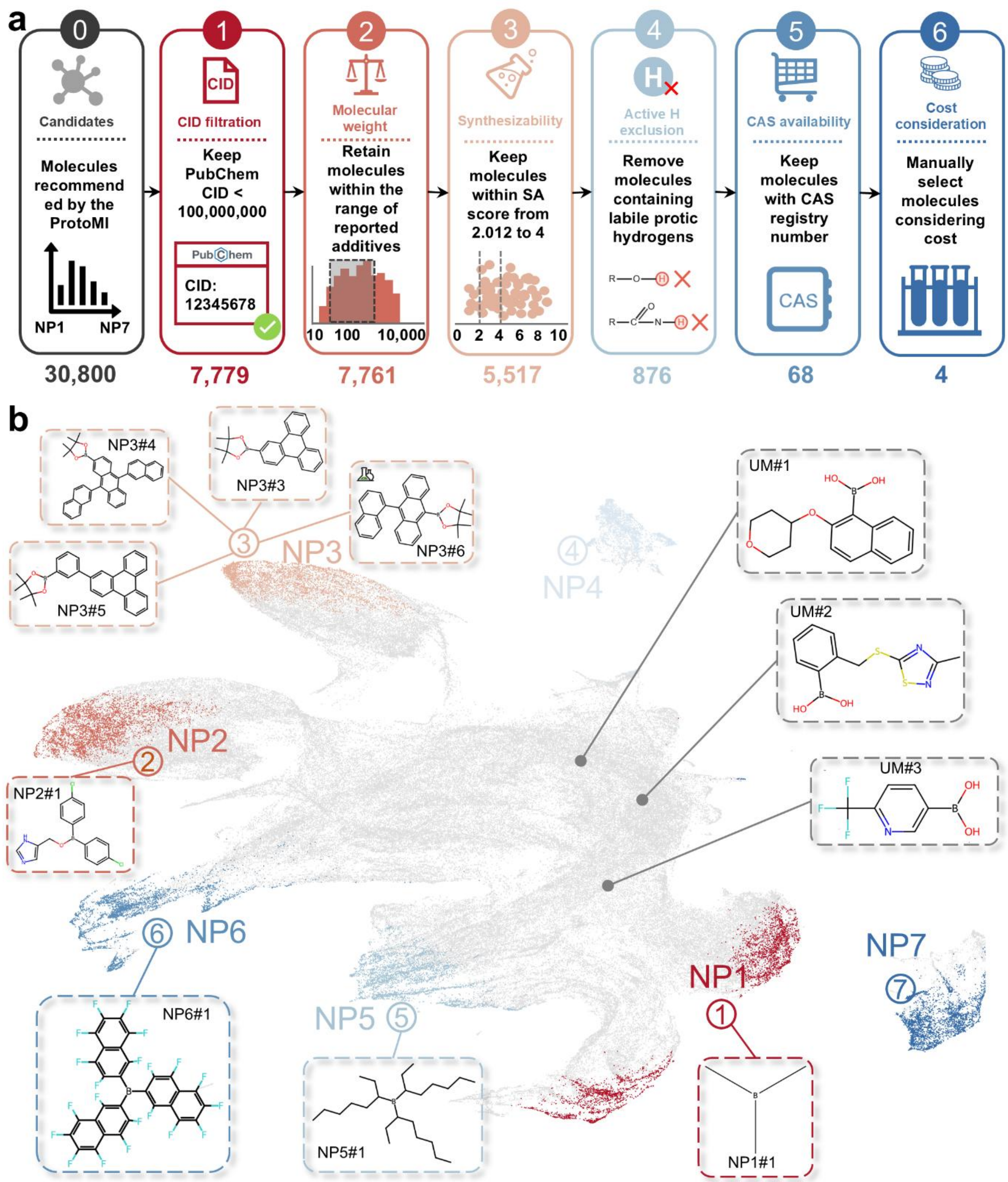

**Fig. 5 | Knowledge-guided translation converts ProtoMI recommendations into experimentally actionable candidates. a,** Post-screening workflow applied to molecules recommended by ProtoMI. Sequential filtering based on PubChem CID validity, molecular-weight range, synthetic accessibility, active hydrogen exclusion, CAS availability and practical cost considerations progressively reduced 30,800 recommended molecules to four experimentally selected candidates. Numbers below each step indicate the remaining candidate pool after filtering. **b,** Distribution of recommended molecules in the transferred molecular representation space. Colored regions denote 7 new prototype domains (NP1-NP7) formed after transferring prototype knowledge to the unlabeled boron-containing molecular space. Representative molecules associated with each new prototype are shown to illustrate the chemical motifs captured by the transferred prototype regions. Representative unlabeled molecules (UM#1-UM#3) are additionally shown from the broader molecular space to indicate structurally plausible molecules that were present in the unlabeled pool but not retained as final experimental candidates.

### Experimental validation in high-temperature $LiFePO_4$||graphite cells

To validate whether a ProtoMI-recommended molecule could translate into measurable battery performance improvement, we selected one representative candidate, 4,4,5,5-tetramethyl-2-[10-(1-naphthyl)anthracen-9-yl]-1,3,2-dioxaborolane (TNDB; NP3#6), for electrochemical testing (Fig. 6a). $LiFePO_4$||graphite cells were chosen because this chemistry is widely used in electric vehicles and stationary energy storage, while its cycling stability at elevated temperature remains limited by parasitic interfacial reactions and transition-metal crossover[32,33]. The additive-containing electrolyte was prepared by introducing 1 wt% TNDB into the baseline electrolyte (1M $LiPF_6$ in ethylene carbonate (EC):dimethyl carbonate (DMC) = 3:7 v/v) and tested at 55 °C. Compared with the baseline electrolyte, the TNDB-containing cell showed improved capacity retention, with a 34.93% improvement at the $100^{th}$ cycle (Fig. 6b). Because high-temperature degradation in $LiFePO_4$||graphite cells is associated with electrolyte decomposition, SEI instability and Fe dissolution/deposition[32,33], we next investigated how TNDB alters the interphase chemistry.

The chemical contribution of TNDB to SEI formation was first examined using X-ray photoelectron spectroscopy (XPS). High-resolution B 1s spectra collected from graphite electrodes recovered from TNDB-containing cells showed B–F and B–O coordination environments after both formation and 100 cycles (Fig. 6c), indicating that TNDB-derived boron species are incorporated into the graphite interphase. XPS after formation further showed a small but detectable B signal only in the TNDB-containing cell, together with a higher F atomic fraction and a lower O atomic fraction relative to the baseline cell (Fig. 6d). These surface-composition changes suggest that TNDB contributes to a B/F-containing interphase and suppresses the accumulation of oxygen-rich decomposition products during early SEI formation.

F 1s XPS was then used to analyze fluorinated interphase species after formation (Fig. 6e). In both electrolytes, LiF remained the dominant fluorinated component, while the TNDB-containing electrolyte showed a modified distribution of LiF, $Li_xPF_y$ and $Li_xPOF_y$ species. Specifically, the LiF contribution remained high in the TNDB-containing cell, whereas the $Li_xPOF_y$ contribution increased relative to the baseline electrolyte. This indicates that TNDB does not simply increase the total LiF fraction at the outermost surface, but instead modifies the formation of F- and P/O-containing inorganic interphase species. Together with the B 1s results, these XPS signatures support the formation of a TNDB-modified inorganic interphase on graphite.

The organic and inorganic carbonate components of the graphite SEI were further analyzed by C 1s XPS peak deconvolution (Fig. 6f and Supplementary Fig. S11). Following common assignments for graphite SEI analysis, the C 1s spectra were fitted into C-C, C-O, C=O, $ROCO_2Li$ and $Li_2CO_3$ components[56,57]. In the baseline electrolyte, oxygenated organic SEI species, including C-O, C=O and $ROCO_2Li$, accounted for 38% of the fitted C 1s area after formation and remained at 32% after 100 cycles. In contrast, the TNDB-containing electrolyte showed lower oxygenated organic contributions, decreasing from 28% after formation to 26% after 100 cycles. The $Li_2CO_3$ contribution after 100 cycles was also reduced from 4% in the baseline cell to 1% in the TNDB-containing cell. These results indicate that TNDB suppresses the accumulation of carbonate-derived organic species and inorganic carbonate residues during high-temperature cycling.

Time-of-flight secondary ion mass spectrometry (ToF-SIMS) provided depth-resolved evidence for TNDB-mediated SEI reconstruction (Fig. 6g and Supplementary Fig. S12). Representative negative secondary-ion fragments, including $CH_3CO^-$, $PO_2^-$, $CO_3^-$ and $LiF_2^-$, were used to visualize solvent-derived organic fragments, $LiPF_6$-derived P-O species, carbonate-containing residues and LiF-related inorganic species, respectively[58,59]. In the baseline electrolyte, $CH_3CO^-$ and $CO_3^-$ fragments extended broadly through the analyzed interphase volume, consistent with heterogeneous accumulation of carbonate-derived decomposition products. In contrast, graphite electrodes cycled with TNDB showed attenuated oxygenated organic fragments and a reorganized distribution of $LiF_2^-$ and $PO_2^-$ species. These depth-resolved signatures are consistent with a graphite interphase that is less dominated by organic carbonate decomposition and contains redistributed salt-derived inorganic components[60,61].

Additive-derived boron incorporation was further visualized using x-z cross-sectional ToF-SIMS maps of the $B^-$ fragment in graphite electrodes cycled with TNDB (Supplementary Fig. S13). A sparse and discontinuous $B^-$ signal was observed after formation, whereas a more extensive $B^-$ distribution appeared after 100 cycles, indicating progressive incorporation and retention of boron-containing species within the SEI. Sputter-depth profiles of representative SEI fragments further showed that $CH_3CO^-$ signals were attenuated in the TNDB-containing electrolyte, supporting suppressed accumulation of solvent-derived organic decomposition products (Supplementary Fig. S14). The $FeO^-$ depth profile was also strongly suppressed after formation and remained lower after extended cycling in the TNDB-containing cell (Supplementary Fig. S15). Because transition-metal deposits on graphite can catalyze electrolyte reduction and destabilize the anode interphase[62], the reduced $FeO^-$ signal suggests that TNDB-derived interphase chemistry mitigates Fe contamination of the graphite surface.

SEM and EDS analyses provided complementary information on surface morphology and Fe deposition. SEM images show the morphology of graphite electrodes recovered after formation and after 100 cycles with and without TNDB (Supplementary Fig. S16). EDS mapping further revealed pronounced Fe accumulation on graphite electrodes cycled in the baseline electrolyte, with the Fe fraction in selected mapping regions increasing from 2.06 wt% after formation to 3.61 wt% after 100 cycles (Fig. 6h). In contrast, graphite electrodes cycled with TNDB showed substantially lower Fe accumulation, with corresponding values of 0.23 wt% after formation and 1.04 wt% after 100 cycles. These values correspond to 89% and 71% lower Fe contents, respectively, relative to the baseline electrolyte. The reduced Fe deposition may help suppress Fe-catalyzed electrolyte decomposition and interphase deterioration during high-temperature cycling.

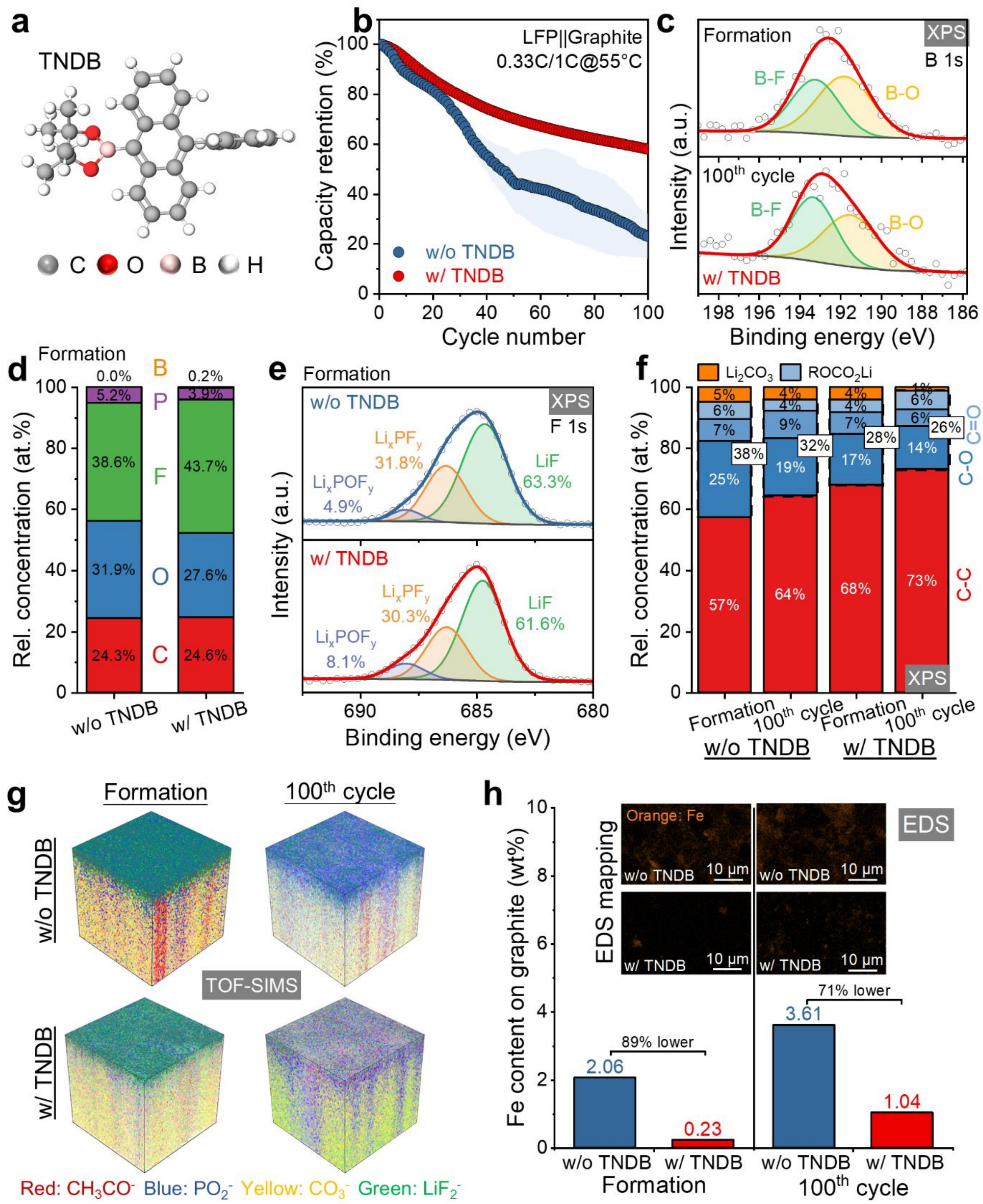

**Fig. 6 | Interphase regulation by the TNDB under 55 °C cycling.** **a**, Molecular structure of TNDB shown as a ball-and-stick model; grey, C; red, O; pink, B; and white, H. **b**, Capacity retention of $LiFePO_4$||graphite coin cells containing the baseline or TNDB-containing electrolyte during cycling at 55 °C. The shadings indicate the deviation of two reproducible cells. **c**, High-resolution B 1s XPS spectra of graphite electrodes recovered from TNDB-containing cells after formation and 100 cycles, with the fitted components assigned to B-F and B-O. **d**, XPS-derived relative atomic concentrations of C, O, F, P and B on graphite electrodes recovered from baseline and TNDB-containing cells after formation. **e**, F 1s XPS spectra of graphite electrodes recovered from cells with and without TNDB after the formation, with the fitted components assigned to LiF, $Li_xPF_y$, $Li_xPOF_y$. **f**, Relative C 1s peak area fractions of C-C, C-O, C-O, $ROCO_2Li$ and $Li_2CO_3$ for graphite electrodes recovered from baseline and TNDB-containing cells after formation and 100 cycles. **g**, Merged three-dimensional ToF-SIMS compositional renderings of $CH_3CO^-$ (red), $PO_2^-$ (blue), $CO_3^-$ (yellow) and $LiF_2^-$ (green) fragments in the corresponding graphite

interphases. **h**, Representative Fe EDS maps (inset) and corresponding semi-quantitative Fe contents of the graphite electrodes.

Operando Fourier transform infrared spectroscopy (FTIR) was collected during the formation step at 55 °C to monitor electrolyte and interphase evolution under operating conditions (Fig. 7a,b). The upper panels show time-resolved absorbance spectra, while the lower panels show difference spectra, $\Delta A(t) = A(t) - A(t_0)$, where $t$ is the time and $t_0$ corresponds to the beginning of charge. In the baseline electrolyte, the difference spectra show a gradual increase in the organic semi-carbonate region. Bands around 1640-1670 $cm^{-1}$ can be assigned to lithium semi-carbonate and lithium alkyl carbonate species, such as lithium methyl carbonate and lithium ethylene dicarbonate, according to previous operando infrared studies[63,64]. Quantitative integration of this region shows a faster growth of organic semi-carbonate species in the baseline cell than in the TNDB-containing cell (Fig. 7c). At the end of charge, the integrated response was 0.25 a.u. $cm^{-1}$ for the baseline electrolyte and 0.17 a.u. $cm^{-1}$ for the TNDB-containing electrolyte, consistent with suppressed formation of carbonate-derived organic SEI species.

The $PF_6^-$-related P-F stretching region also evolved differently in the two electrolytes. We use the integrated $PF_6^-$ response between 820-880 $cm^{-1}$ to track $PF_6^-$-related interfacial evolution. The baseline electrolyte developed a positive end-point response in this region, whereas the TNDB-containing electrolyte showed a negative end-point response (Fig. 7d). This contrast indicates that TNDB changes the direction and magnitude of $PF_6^-$-related spectral evolution during formation. When considered together with the XPS and ToF-SIMS results, the operando FTIR data suggests that TNDB suppresses carbonate-solvent decomposition while modifying salt-derived interfacial chemistry.

Overall, the post-mortem and operando evidence supports a coherent interphase-regulation mechanism for improved high-temperature cycling performance. TNDB participates in graphite interphase construction by introducing B-containing species, modifying F-, P-, and O-containing inorganic components, suppressing carbonate-derived organic SEI accumulation and reducing Fe deposition on graphite. These coupled effects provide a plausible explanation for the improved capacity retention of $LiFePO_4$||graphite cells cycled at 55 °C with the ProtoMI-recommended TNDB additive.

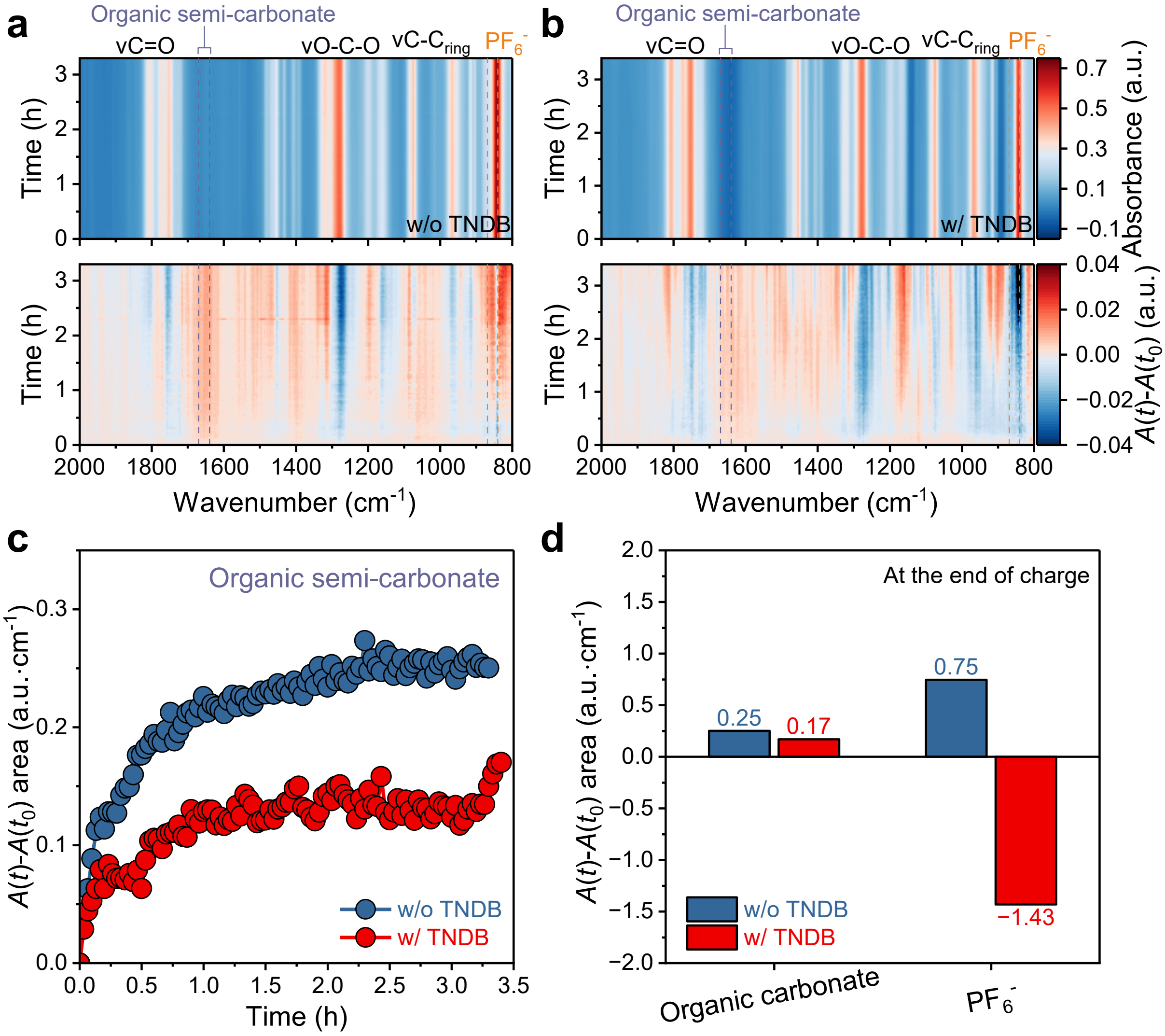

**Fig. 7 | Operando FTIR analysis of electrolyte and interphase evolution during 55 °C formation. a,b,** Time-resolved absorbance spectra (top) and corresponding difference spectra, $A(t)$-$A(t_0)$ (bottom), collected during the formation-charge step at 55 °C for $LiFePO_4$||graphite cells containing the baseline electrolyte (**a**) and the TNDB-containing electrolyte (**b**). The dashed lines delimit the 1640-1670 $cm^{-1}$ region associated with organic semi-carbonate species. The P-F stretching region of $PF_6^-$ is indicated at approximately 820-880 $cm^{-1}$. Identical color scales are used for direct comparison. **c,** Time evolution of the integrated absolute difference absorbance within the organic semi-carbonate region with and without TNDB. **d,** Integrated spectral changes for organic semi-carbonate and the $PF_6^-$ P-F stretching region at the end of the 1$^{st}$ charge.

## Discussion and conclusion

In this work, we develop ProtoMI as a molecular discovery framework for data-scarce settings in which experimentally validated molecules are sparse and most candidate molecules remain unlabeled. Rather than treating additive discovery as a conventional supervised property-prediction problem, ProtoMI re-formulates it as an efficient exploration problem. Sparse literature-derived successes are converted into transferable prototype representations, adapted to a broader unlabeled candidate space and translated into experimentally actionable molecules through practical chemical constraints. This workflow allows limited experimental resources to be concentrated on chemically promising regions rather than distributed across the full candidate space.

A central feature of ProtoMI is that knowledge transfer occurs at the prototype level rather than at the level of individual molecules. Conventional similarity-based recommendation strategies tend to remain close to previously reported compounds and may therefore restrict exploration to local molecular neighborhoods. In contrast, ProtoMI learns higher-order organization from groups of reported additives and transfers this organization across source-target distribution mismatch. The preservation and redistribution of prototype identities after adaptation suggest that the framework captures transferable chemical relationships instead of simply memorizing known molecular structures. This prototype-centric strategy provides a practical route to extending sparse literature knowledge into structurally broader regions of chemical space.

The experimental validation demonstrates how ProtoMI recommendations can be connected to battery performance. As a representative candidate, TNDB improved the high-temperature cycling stability of $LiFePO_4$||graphite cells relative to the baseline electrolyte. XPS, ToF-SIMS, EDS mapping and operando FTIR analyses indicate that TNDB incorporates B-containing species into the graphite interphase, modifies F/P/O-containing inorganic SEI components, suppresses carbonate-solvent decomposition, alters $PF_6^-$-related spectral evolution and reduces Fe deposition on graphite. These coupled interfacial effects provide a plausible explanation for the improved capacity retention observed at 55 °C. At the same time, the present experimental validation should be interpreted as a proof of concept for one representative ProtoMI-recommended candidate rather than exhaustive validation of all recommended molecules.

More broadly, this study highlights the importance of discovery efficiency in scientific problems where labels are scarce, and candidate spaces are large. Under such conditions, maximizing predictive accuracy for individual molecules may be less practical than improving the allocation of limited experimental resources. Retrospective temporal validation shows that literature-derived prototype knowledge can enrich future-reported additives while screening only a small fraction of the candidate space. This suggests that prototype-guided exploration can complement supervised prediction, physics-informed modeling and active-learning workflows by serving as an upstream candidate-prioritization strategy.

Several limitations remain. The current implementation focuses on boron-containing electrolyte additives and relies primarily on structural molecular representations. It does not yet explicitly incorporate electrochemical stability, reaction pathways, solvation structures, synthesis routes or cell-level performance labels. In addition, the translation stage depends on empirical filters and commercial availability constraints, which may bias the final experimentally selected candidates. Future studies could integrate physics-informed descriptors, quantum-chemical calculations, synthesis-planning tools, multimodal molecular representations and active experimental feedback to further improve recommendation quality and practical accessibility. With these extensions, prototype-guided knowledge transfer may provide a general route for molecular discovery under severe label scarcity.

## Methods

**M1. Data collection.** For the searching space molecules dataset, we used keywords (*Borane, Borate, Boric, Boron, Boronic, Boronate, Additive, Electrolyte*) to inquire the boron-related molecules and their molecular information, including formula, SMILES, fingerprint, topological, weight, and heavy atom. For the positive samples dataset, we used the heuristic extraction process to obtain the literature information of each reported boron-containing electrolyte additive molecule. The details of the LLM-assisted data-mining pipeline is provided in Supplementary Fig. S17. Finally, we get 179,977 boron-containing molecules as the searching space molecules dataset and 126 reported boron electrolyte additive molecules as the positive samples dataset.

**M2. Graph-level descriptors.** To characterize the global topological properties of molecules, several graph-level descriptors were calculated, including atom number ($N$), bond number ($E$), average degree ($\bar{k}$), graph density ($D$), and degree entropy ($H$). The average degree was defined as

$$\bar{k} = \frac{1}{N}\sum_{i=1}^{N} k_i = \frac{2E}{N} \tag{1}$$

where $k_i$ denotes the degree of node $i$. Graph density, which measures the fraction of existing bonds relative to the maximum possible number of bonds, was calculated as

$$D = \frac{2E}{N(N-1)} \tag{2}$$

To quantify the heterogeneity of node connectivity, the degree entropy was computed using the Shannon entropy of the degree distribution,

$$H = -\sum_{k} p(k) \log p(k) \tag{3}$$

where $p(k)$ denotes the probability that a node has degree $k$. These descriptors collectively characterize molecular size, connectivity and topological complexity.

**M3. Unsupervised graph contrastive learning (UCL) model.** To extract prototype knowledge from experimentally reported boron electrolyte additive molecules, an unsupervised graph contrastive learning strategy was first applied to all reported boron electrolyte additive molecules. Each molecule was represented as a graph $RG_i = (RV_i, RE_i)$, where node features describe atomic attributes and edge features describe bond-level information. During training, two views of each molecular graph were constructed for contrastive learning. The first view preserves the original structure, while the second view is generated via stochastic edge perturbation:

$$RG_i^{(1)} = RG_i, \tag{4}$$

$$RG_i^{(2)} = \left(RV_i, \left(RE_i \setminus RE_i^{del}\right) \cup RE_i^{add}\right), \tag{5}$$

$$\left|RE_i^{del}\right| = \left|RE_i^{add}\right| = \left\lfloor \frac{\rho |RE_i|}{2} \right\rfloor \tag{6}$$

where $RE_i^{del}$ and $RE_i^{add}$ denote the sets of removed and newly added edges, respectively, and $\rho = 0.1$ controls the perturbation ratio. The two graph views were then encoded by a three-layer graph neural network (GNN). The layer-wise propagation can be formulated as:

$$H^{(l)} = Dropout\left(ReLU\left(GraphNorm\left(GINE^{(l)}\left(H^{(l-1)}, RE_i\right)\right)\right)\right), l = 1,2,3 \tag{7}$$

where $H^{(0)} = RX_i$ denotes the initial node feature matrix and $l$ represents the layer number of this GNN encoder. The graph-level representation was obtained via global mean pooling:

$$h_{RG_i} = f_\theta^{USL}(RG_i) = MeanPool\left(H^{(3)}\right) = \frac{1}{RV_i} \sum_{v \in RV_i} h_v^{(3)} \tag{8}$$

where $h_v^{(3)}$ represents the embedding of node after the third layer GNN, and $f_\theta^{USL}$ is the three-layer encoder for this unsupervised learning stage. To optimize the representation space, the graph embeddings were further mapped by projection heads $g_1(\cdot)$and $g_2(\cdot)$:

$$z_i^{(1)} = g_1\left(h_{RG_i}^{(1)}\right), z_i^{(2)} = g_2\left(h_{RG_i}^{(2)}\right) \tag{9}$$

The use of separate projection heads follows standard contrastive learning practice to improve representation quality. The model was trained using an InfoNCE contrastive loss (Method M10), which encourages embeddings from two augmented views of the same molecule to be close, while separating them from other molecules in the batch. After each training epoch, molecular embeddings were L2-normalized and clustered using hierarchical clustering with outlier filtering, average linkage and cosine distance (Method M9). The optimal number of clusters was determined by maximizing the average silhouette score (Method M5) over a predefined range. The model achieving the highest silhouette score was selected as the final representation model and used to define molecular prototypes for subsequent prototype-guided learning.

**M4. Prototype semi-supervised contrastive learning model.** To transfer prototype knowledge from experimentally reported boron electrolyte additive molecules to the unlabeled molecular space, we developed a prototype semi-supervised contrastive learning model. First, we use the pre-trained molecular representation model (Method M3) to generate prototype-level embeddings for all experimentally reported additive molecules $h_{RG_i}$, and followed by a randomly initialized projection head which is motivated by the need to improve robustness and avoid representation bias induced by the contrastive pretraining objective:

$$\boldsymbol{z}_i = Norm\left(g'\left(f_\theta^{USL}(RG_i)\right)\right) \tag{10}$$

where $g'(\cdot)$ denotes the projection head. Hierarchical clustering with average linkage and cosine distance was then applied to the normalized positive-sample embeddings. The optimal number of clusters was determined by comparing clustering quality across candidate cluster numbers. Each cluster was regarded as a molecular prototype group, and the corresponding prototype label table $RL$ was assigned to each positive additive:

$$RL = \mathrm{HierarchicalCluster}(\boldsymbol{z}_i) \tag{11}$$

Next, positive additives and unlabeled candidate molecules were jointly used to train the prototype semi-supervised contrastive learning model. At each epoch, prototype centroids were first recalculated from the positive additives according to their prototype labels. For prototype $n$, the centroid was computed as the mean representation of positive samples assigned to this prototype:

$$p_{n,new}^{(t)} = Norm\left(\frac{1}{|\mathcal{P}_n|}\sum_{i\in\mathcal{P}_n} \boldsymbol{z}_i^{(t)}\right) \tag{12}$$

where $\mathcal{P}_n$ denotes the set of positive additives assigned to prototype $n$, and $\boldsymbol{z}_i^{(t)}$ is the normalized embedding produced by the current model at epoch $t$. To avoid unstable prototype shifts during semi-supervised adaptation, the prototype centroids were updated using an exponential moving average[55]:

$$p_n{}^{(t)} = mp_n^{(t-1)} + (1-m)p_{n,new}^{(t)}, m = 0.999 \tag{13}$$

where $m$ is the momentum coefficient. This update allows the prototypes to gradually adapt from the positive-only representation space toward the broader unlabeled molecular space while preserving the original prototype knowledge extracted from experimentally reported additives.

At the beginning of the model training, a new GNN encoder and projection head were initialized and trained using both positive and unlabeled molecular graphs. For each mini-batch, the similarity between molecular embeddings and prototype centroids was calculated by cosine similarity. For each prototype, the top-k molecules with the highest similarity to that prototype were selected as prototype-relevant core samples. The embeddings of the selected molecules can be expressed in:

$$\boldsymbol{z}_i = \frac{g''\left(f_\theta^{PCL}(G_i)\right)}{\left\|g''\left(f_\theta^{PCL}(G_i)\right)\right\|_2}, \tag{14}$$

$$s_{in} = sim(\boldsymbol{z}_i, p_n), \tag{15}$$

$$\mathcal{I}_n^{(K)} = \mathrm{TopK}(\{s_{1n}, s_{2n}, \dots, s_{Nn}\}), \tag{16}$$

$$\mathcal{Z}_n = \left\{\mathbf{z}_i \mid i \in \mathcal{I}_n^{(K)}\right\}. \tag{17}$$

where $G_i \in \{RG_i, UG_i\}$ denotes either a reported positive molecular graph or an unlabeled molecular graph; $f_\theta^{PCL}$ is the newly initialized GNN encoder; $g''(\cdot)$ is the newly initialized projection head; $\mathbf{z}_i$ is the normalized embedding of molecule $i$; $p_n$ denotes the centroid of the $n$-th prototype; $s_{in}$ is the cosine similarity between molecule $i$ and prototype $n$; $N$ is the number of molecules in the mini-batch; $\mathcal{I}_n^{(K)}$ represents the index set of the top-k molecules most similar to prototype $n$; and $\mathcal{Z}_n$ denotes the corresponding set of selected prototype-relevant embeddings. In this work, $K = 25$.

This strategy restricts contrastive optimization to molecules that are most relevant to each prototype, thereby reducing the influence of low-confidence unlabeled samples and preventing noisy candidates from dominating the representation learning process. For prototype $n$, the selected molecules were optimized using a prototype contrastive loss (Method M10). To encourage different prototypes to encode distinct molecular patterns, a prototype decorrelation regularization term was introduced (Method M10). The model was optimized using Adam with weight decay. After each epoch, the learned embeddings were evaluated on the held-out positive and unlabeled test samples. The silhouette score (Method M5) based on cosine distance was used to quantify the separation quality of the learned prototype-aware representation space. The model checkpoint with the highest silhouette score was selected as the best prototype semi-supervised contrastive learning model.

**M5. Evaluating metrics.** The silhouette score was used to assess the quality of the learned embedding space and the separation of prototype clusters. For each sample, the silhouette coefficient compares its average intra-cluster distance $a(i)$ with the minimum average distance to the nearest neighboring cluster $b(i)$, defined as:

$$Silhoutte\ Score = \frac{1}{N}\sum_{i=1}^{N}\frac{b(i)-a(i)}{\max\big(a(i),\ b(i)\big)} \tag{18}$$

The score ranges from −1 to 1, where higher values indicate more compact clusters and better inter-cluster separation. In this study, the silhouette score was computed using cosine distance for consistency with the clustering metric. The optimal number of clusters was selected by maximizing the average silhouette score. Furthermore, it was used as an indicator of representation quality, where higher scores reflect more discriminative and well-structured molecular embeddings.

To evaluate both prototype similarity and prototype coverage, an entropy-weighted prototype similarity (EWPS) metric was defined. Each recommended molecule was assigned to the prototype centroid with the highest cosine similarity, and the average maximum similarity was calculated as:

$$\bar{s} = \frac{1}{N}\sum_{i=1}^{N} s_i \tag{19}$$

where $s_i$ denotes the maximum cosine similarity between molecule $i$ and all prototype centroids. Prototype coverage was quantified using the normalized assignment entropy:

$$H_{norm} = \frac{-\sum_{k=1}^{K} p_k ln p_k}{ln\ K} \tag{20}$$

where $p_k$ is the fraction of molecules assigned to prototype $k$. The final EWPS metric was defined as:

$$EWPS = \bar{s} * H_{norm} \tag{21}$$

Higher EWPS values indicate that the recommended molecules are both highly representative of the learned prototypes and broadly distributed across prototype categories.

To evaluate the ability of ProtoMI to recover future-reported additives, a temporal validation strategy was adopted. Recall was defined as:

$$Recall = \frac{N_{hit}}{N_{hidden}} \tag{22}$$

where $N_{\mathrm{hit}}$ is the number of hidden positive molecules recovered by the recommendation list and $N_{\mathrm{hidden}}$ is the total number of hidden positives. Higher recall values indicate a stronger capability for identifying future experimentally reported additives.

To quantify the efficiency improvement over random screening, the enrichment factor (EF) was calculated as:

$$EF = \frac{N_{hit}/N_{rec}}{N_{hidden}/N_{candidate}} \tag{23}$$

where $N_{\mathrm{rec}}$ is the number of recommended molecules and $N_{\mathrm{candidate}}$ is the size of the candidate pool. EF measures how many times the recommendation strategy outperforms random selection, with larger values indicating more efficient prioritization of experimentally promising molecules.

**M6. Translation strategy.** To further improve experimental feasibility, we designed a knowledge-guided post-screening pipeline based on empirical chemical constraints, as illustrated in Fig. 5a. Specifically, six sequential filtering criteria were applied. First, CID filtration was performed by retaining only molecules with PubChem compound IDs (CID) smaller than 100,000,000, as larger IDs typically correspond to composite entries rather than single molecular entities. Second, molecular weight filtering was conducted by constraining candidates within the empirical range observed in reported boron-containing additives (26.811–1455.489). Third, synthesizability was evaluated using the SA score, and only molecules within the dataset-derived range (2.012–6.319) were preserved. Fourth, active hydrogen exclusion was implemented via SMiles

ARbitrary Target Specification (SMARTS)-based functional group matching[65] to remove molecules containing labile protic hydrogens, which are known to negatively affect electrolyte stability and battery performance. Fifth, CAS availability was required to ensure that selected molecules are commercially accessible. Finally, considering practical cost constraints, we manually selected four molecules from 68 commercially available candidates for experimental validation.

**M7. Molecular representation.** In this paper, we used fingerprints, SMILES, and graphs to represent molecules and used the augmented molecular graph representation (Method M8) as the best performance molecular representation method for the model training stage. For fingerprints, we used the *rdFingerprintGenerator.GetMorganGenerator()* function in the RDKit Python package to generate fingerprints for all molecules with *radius=2* and *fpSize=1024*. For SMILES, most SMILES can be obtained from PubChem, OPSIN, and the literature, while some molecules are lacking, and we used ChemDraw software to generate them manually. For graphs, we converted the SMILES into the molecular graph by encoding atom elements, the number of heavy neighbors, formal charge, hybridization types, whether it is in a ring, whether it is aromatic, scaled van der Waals radius, and scaled covalent radius information as the node features, and bond types, bond is conjugated, bond is in ring, bond direction, and bond is aromatic information as the edge features. All numeric features have been normalized to the float data type, and all string features are one-hot encoded. See the Supplementary Tables S2 and S10 for details.

**M8. Graph data augmentation.** To enhance the robustness of representation learning under limited labeled data, we employ a multi-scale graph augmentation strategy that perturbs molecular graphs at the feature, structure, and topology levels. At the feature level, node attributes are augmented through random feature shuffling[42] and stochastic noise masking[39]. Feature shuffling randomly permutes a subset of node feature dimensions across nodes, while noise masking injects Gaussian perturbations into randomly selected feature entries. These operations simulate uncertainty and variability in molecular descriptors. At the structural level, graph topology is modified through node dropping[43] and edge perturbation[40]. Node dropping randomly removes a subset of nodes and their associated edges, generating subgraph-level variations. Edge perturbation is performed using a weighted sampling strategy, where edges are probabilistically removed or added based on node degree and shortest-path distance, encouraging the model to learn robust local structural patterns. At the global topology level, we further introduce a Personalized PageRank (PPR)-based augmentation[41]. A PPR diffusion matrix is computed to capture higher-order connectivity, and edges are selectively removed or added according to their PPR scores. In particular, the diffusion matrix is computed to capture higher-order connectivity, defined as:

$$P = \alpha \sum_{k=0}^{K} (1-\alpha)^k \hat{A}^k \tag{24}$$

where $\hat{A} = D^{-\frac{1}{2}} A D^{-\frac{1}{2}}$ is the normalized adjacency matrix, $D$ is the degree matrix with $D_{ii} = \sum_j A_{ij}$, $A$ is the adjacency matrix encoding the bonding topology of the molecular graph, $\alpha$ is the teleport probability, and $K$ is the truncation depth. The diffusion score $P_{ij}$ quantifies higher-order topological relationships between node $i$ and node $j$. Based on this score, edges with lower $P_{ij}$ values are preferentially removed, while non-edge pairs with higher $P_{ij}$ values are more likely to be added.

**M9. Hierarchical clustering algorithm with outlier filtering.** To identify representative prototype groups from experimentally reported positive additives, hierarchical clustering was performed on the learned molecular graph embeddings using average linkage with cosine distance. Different clustering solutions were generated by varying the number of clusters from 3 to a predefined maximum value, *max_cluster=10*, and the optimal cluster number (7) was selected according to the silhouette score (Method M5) calculated with cosine distance. To improve the robustness of prototype discovery, a cluster-wise outlier filtering strategy was applied for each candidate clustering solution. For each cluster, the centroid was calculated as the mean embedding of all samples assigned to that cluster. The cosine similarity between each sample embedding and its corresponding cluster centroid was then computed and converted into cosine distance as $1 - \cos(x_i, c_k)$, $x_i$ represents the $i$th sample, and $c_k$represents the $k$th prototype centroid. Samples with distances larger than the cluster-specific threshold, defined as the mean distance plus 1.5 times the standard deviation, were regarded as outliers and removed. Clusters containing fewer than two samples were excluded from this filtering procedure. After outlier removal, hierarchical clustering was recomputed on the filtered embeddings using average linkage and cosine distance. The silhouette score was then evaluated for each candidate cluster number, and the clustering configuration with the highest score was selected. The corresponding filtered embeddings, cluster labels, additive names, and linkage matrix were retained for subsequent prototype construction and prototype-guided representation learning.

**M10. Loss functions.** In this section, we introduce two loss functions used in the pre-trained molecular representation model and prototype semi-supervised contrastive learning model, respectively.

For the pre-trained molecular representation model, an InfoNCE-based contrastive loss was adopted to learn discriminative molecular representations. Given two augmented views of each molecular graph, their embeddings were first L2-normalized, and cosine similarity was used to measure pairwise similarity. For each sample, the corresponding positive pair was encouraged to have higher similarity than all other samples in the batch, which served as negatives. The loss is defined as:

$$\mathcal{L}_{InfoNCE} = -\frac{1}{N}\sum_{i=1}^{N} log \frac{\exp\left(sim\left(z_i^{(1)}, z_i^{(2)}\right)/\tau\right)}{\sum_{j=1}^{N}\exp\left(sim\left(z_i^{(1)}, z_j^{(2)}\right)/\tau\right)} \tag{25}$$

where $\tau$ is a temperature parameter controlling the concentration level of the distribution, and $z$ represents the embedding of each sample. All other samples in the batch are treated as negatives. The similarity function $\mathrm{sim}(\cdot,\cdot)$ is defined as the cosine similarity between two embedding vectors:

$$sim\left(z_i^{(1)}, z_i^{(2)}\right) = \frac{{z_i^{(1)}}^{\top} z_i^{(2)}}{\left\|z_i^{(1)}\right\|\left\|z_i^{(2)}\right\|} = {z_i^{(1)}}^{\top} z_i^{(2)} \tag{26}$$

For the prototype semi-supervised contrastive learning model, the similarity between a selected molecule and its corresponding prototype was treated as the positive logit, while similarities to all other prototypes were treated as competing logits:

$$\mathcal{L}_{proto} = -\frac{1}{M}\sum_{m=1}^{M}\frac{1}{|\mathcal{Z}_m|}\sum_{\boldsymbol{z}\epsilon\mathcal{Z}_m} log \frac{\exp(sim(\boldsymbol{z}, p_m)/\tau)}{\sum_{r=1}^{M}\exp(sim(\boldsymbol{z}, p_r)/\tau)} \tag{27}$$

where $M$ is the number of prototypes, and $\tau = 0.1$ is the temperature parameter, and, $\mathcal{Z}_m$ was constructed by top-k assignment within each batch, such that molecules with the highest similarity to $p_m$ were treated as positive samples for that prototype.

$$\mathcal{L}_{decor} = \frac{1}{M^2}\sum_{i=1}^{M}\sum_{j=1}^{M}\left(sim(p_i, p_j) - \delta_{ij}\right)^2, \delta_{ij} = 1\ if\ i = j, else\ 0 \tag{28}$$

where $p_i \in \mathbb{R}^d$represents the embedding vector of the $i$-th prototype centroid in the latent space; ${p_i}^{\top}p_j$ denotes the inner product (cosine similarity after normalization) between prototype $i$ and prototype $j$; and $\delta_{ij}$ is the Kronecker delta, which equals 1 when $i = j$ and 0 otherwise. This encourages prototypes to capture diverse and non-overlapping molecular patterns. The total loss was defined as the sum of the prototype contrastive loss and the prototype decorrelation loss:

$$\mathcal{L}_{total} = \mathcal{L}_{proto} + \mathcal{L}_{decor} \quad (29)$$

During evaluation, prototype-relevant core samples were selected from the held-out positive and unlabeled test molecules using the same top-k criterion, and the silhouette score with cosine distance was calculated on these selected embeddings.

**M11. Electrochemical measurements.** Electrochemical measurements were carried out using CR2025-type $LiFePO_4$||graphite coin cells. The baseline electrolyte consisted of 1 M $LiPF_6$ in EC:DMC (3:7 by volume), with a solvent water content below 50 ppm, and the additive-containing electrolyte was prepared by adding 1 wt% additive to the baseline electrolyte. The cells were assembled in an Ar-filled glovebox using $LiFePO_4$ cathodes, graphite anodes, Celgard 2025 separators and 60 μL electrolyte, followed by resting for 12 h before testing. The cells were cycled between 2.5 and 3.65 V at 55 °C. Formation was conducted for three cycles at 0.33C charge and 0.33C discharge, followed by long-term cycling at 0.33C charge and 1C discharge. Capacity retention was calculated by normalizing the discharge capacity of each cycle to the second discharge capacity.

**M12. Post-mortem sample preparation.** The coin cells were disassembled in an Ar-filled glovebox to avoid exposure to air and moisture. The graphite electrodes were carefully collected and rinsed with anhydrous DMC three times to remove residual electrolyte and loosely attached salts. After the residual DMC evaporated inside the glovebox, the electrodes were mounted on the corresponding sample holders and transferred to the characterization chambers using an air-free vacuum transfer stage.

**M13. X-ray photoelectron spectroscopy.** XPS was performed using a PHI VersaProbe 4 system (ULVAC-PHI). Survey spectra and high-resolution spectra were collected to analyze the surface chemical composition and bonding environments of the SEI. The C 1s, O 1s, F 1s, P 2p, B 1s, and Fe 2p regions were analyzed when applicable. The binding energies were calibrated using the C-C component in the C 1s spectrum at 284.8 eV. Peak deconvolution was performed after background subtraction to determine the relative contents of different SEI components.

**M14. Time-of-flight secondary ion mass spectrometry.** ToF-SIMS measurements were carried out using an IONTOF M6 instrument. Negative-ion mode was mainly used to analyze SEI-related fragments, including $PO_2^-$, $CH_3CO^-$, $LiF_2^-$, $CO_3^-$, $CH_3COO^-$, and $FeO^-$. Three-dimensional chemical renderings and sputter-depth profiles were obtained by sequential ion analysis and sputtering. The depth profiles were used to compare the relative distribution of solvent-derived organic fragments, carbonate-containing residues, LiF-related species, phosphate/fluorophosphate-related fragments, and Fe-containing deposits within the SEI.

**M15. Scanning electron microscopy.** SEM was conducted using an SU3900 microscope. SEM images were collected to examine the morphology of the cycled graphite electrodes and the surface evolution after high-temperature cycling.

**M16. Energy-dispersive X-ray spectroscopy mapping.** EDS mapping was performed on the cycled graphite electrodes to analyze the spatial distribution of Fe. Elemental mapping was conducted on representative regions of the graphite electrode surface to evaluate transition-metal deposition after formation and long-term cycling. The Fe mapping results were used to compare the extent of Fe accumulation on graphite electrodes cycled in the baseline and additive-containing electrolytes.

**M17. Operando IR spectroscopy collection.**

Operando infrared measurements were conducted using a Bruker Invenio R spectrometer equipped with a customized electrochemical cell containing an embedded IR fiber. The manually polished fiber was inserted through symmetric through-holes and sealed with epoxy resin for 12 h. The cell was assembled by sequentially stacking a graphite electrode, the IR fiber, a Whatman GF/D separator wetted with 120 μL electrolyte, and an LFP electrode with a diameter of 12 mm within a PTFE stabilizing ring. After a 1 h rest, transmitted IR signals were collected by a liquid-nitrogen-cooled MCT detector over 5000-600 $cm^{-1}$. Spectra were acquired every 1 min during cycling at a resolution of 4 $cm^{-1}$ with 32 scans, while maintaining the signal intensity at approximately 10,000 to ensure a high signal-to-noise ratio.

## Acknowledgments

The authors are grateful for the National Natural Science Foundation of China (No. 92372109), the National Key R&D Program of China (No. 2023YFB2503600), and Guangdong S&T Program (No. 2025A0505000017). The authors thank the Materials, Design and Manufacturing Facility (MDMF), Materials Characterization and Preparation Facility (MCPF), and Wilson Tang Brilliant Energy Science and Technology Lab (BEST Lab) of The Hong Kong University of Science and Technology (Guangzhou) for providing experimental support. This work is partially supported by GDIC.

## Author contributions

W.H. conceived the project, designed the methodology, collected the data, developed the model, performed the data analysis, prepared the figures, and wrote the paper, with participation in electrochemical testing and materials characterization. H.D. performed all electrochemical measurements, materials characterization experiments, and wrote the experimental part of the paper. J.T. contributed to the conceptualization of the project. R.T. provided guidance on model development and algorithm design. J.L. supervised the computational work and contributed to paper revision. Y.Q. supervised the paper preparation and provided theoretical guidance on boron chemistry. J.H. conceived and supervised the overall project and contributed to refinement of the paper and figures. All authors discussed the results and contributed to the final paper.

## Competing interests

A Chinese patent application related to this work has been filed and is currently pending.

## Data and code availability

The source code used in this study is publicly available at https://github.com/HongWxd/ProtoMI. The datasets generated and analyzed during the current study are available from the corresponding author upon reasonable request.

# Supplementary information

# Prototype-guided transfer of sparse literature knowledge for electrolyte additive discovery

Weixiang HONG[1,5], Hongting DU[1,5], Jiayue TANG[1], Ruifeng TAN[1], Yangjian QUAN[2], Jia LI[3,*], Jiaqiang HUANG[1,4,*]

[1] Sustainable Energy and Environment Thrust and Guangzhou Municipal Key Laboratory of Materials Informatics, The Hong Kong University of Science and Technology (Guangzhou), Nansha, Guangzhou, 511400, Guangdong, P.R. China

[2] Department of Chemistry, The Hong Kong University of Science and Technology, Clear Water Bay, Kowloon, Hong Kong SAR, P.R. China

[3] Data Science and Analytics Thrust and Guangzhou Municipal Key Laboratory of Materials Informatics, The Hong Kong University of Science and Technology (Guangzhou), Nansha, Guangzhou, 511400, Guangdong, P.R. China

[4] Academy of Interdisciplinary Studies, The Hong Kong University of Science and Technology, Clear Water Bay, Kowloon, Hong Kong SAR, P.R. China

[5] These authors contributed equally: Weixiang HONG, Hongting DU

* Corresponding authors:

J. Huang: seejhuang@hkust-gz.edu.cn

J. Li: jialee@hkust-gz.edu.cn

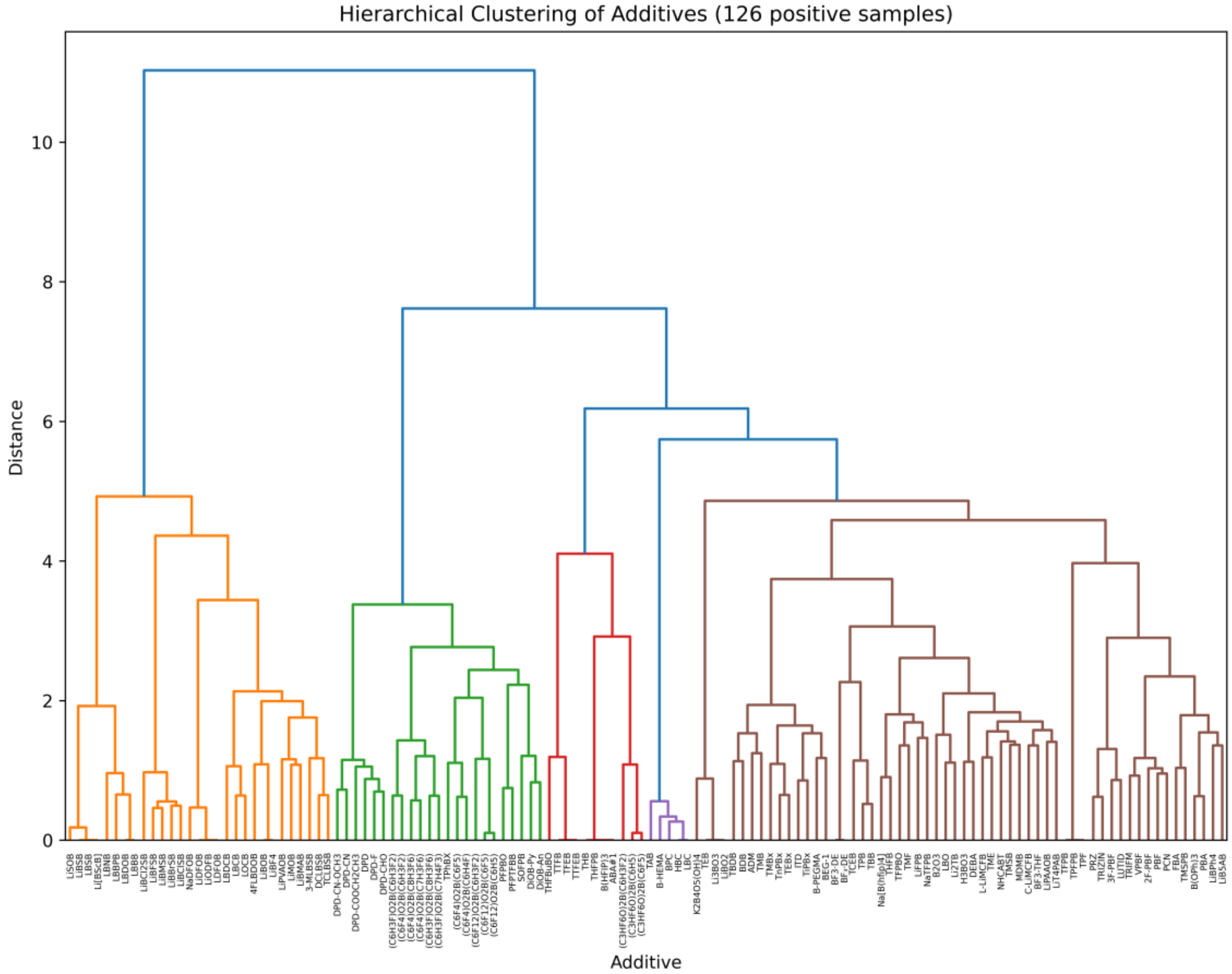

**Figure S1 | Hierarchical clustering[1] dendrogram of reported boron-containing electrolyte additives.** Hierarchical clustering dendrogram of 126 reported boron-containing electrolyte additives constructed using Morgan fingerprints and pairwise Tanimoto distances[2]. Each terminal leaf represents one reported additive, and the branch height reflects molecular dissimilarity. Even with this simple fingerprint-based representation, additives with highly similar names and closely related chemical structures are grouped into the same local branches, such as families of lithium borates, fluorinated borates and aryl borate derivatives. This branch-level organization suggests that reported boron-containing additives are not randomly scattered as isolated successful examples, but are enriched in recurrent structural families. These repeated local branches therefore support the presence of recurring molecular patterns or scaffold-level motifs in the reported additive chemical space. Branch colors are used to visually distinguish major dendrogram regions generated by hierarchical clustering.

**Table S1 | Dataset statistics**

| Category | Value |
| --- | --- |
| Atoms | 41.191±47.281 |
| Bonds | 87.259±102.386 |
| Average degree | 4.130±0.432 |
| Graph density | 0.189±0.117 |
| Degree entropy | 3.265±0.802 |

These statistics were calculated from both reported molecules and the searching space molecules.

**Table S2 | Node features and corresponding physical meanings**

| Feature No. | Physical Meaning | Feature type | Encoding type | Data type |
|---|---|---|---|---|
| N0 | C | Node feature | One hot encoding | String |
| N1 | N | Node feature | One hot encoding | String |
| N2 | O | Node feature | One hot encoding | String |
| N3 | S | Node feature | One hot encoding | String |
| N4 | F | Node feature | One hot encoding | String |
| N5 | Si | Node feature | One hot encoding | String |
| N6 | P | Node feature | One hot encoding | String |
| N7 | Cl | Node feature | One hot encoding | String |
| N8 | Br | Node feature | One hot encoding | String |
| N9 | Mg | Node feature | One hot encoding | String |
| N10 | Na | Node feature | One hot encoding | String |
| N11 | Ca | Node feature | One hot encoding | String |
| N12 | Fe | Node feature | One hot encoding | String |
| N13 | As | Node feature | One hot encoding | String |
| N14 | Al | Node feature | One hot encoding | String |
| N15 | I | Node feature | One hot encoding | String |
| N16 | B | Node feature | One hot encoding | String |
| N17 | V | Node feature | One hot encoding | String |
| N18 | K | Node feature | One hot encoding | String |
| N19 | Tl | Node feature | One hot encoding | String |
| N20 | Yb | Node feature | One hot encoding | String |
| N21 | Sb | Node feature | One hot encoding | String |
| N22 | Sn | Node feature | One hot encoding | String |
| N23 | Ag | Node feature | One hot encoding | String |
| N24 | Pd | Node feature | One hot encoding | String |
| N25 | Co | Node feature | One hot encoding | String |
| N26 | Se | Node feature | One hot encoding | String |
| N27 | Ti | Node feature | One hot encoding | String |
| N28 | Zn | Node feature | One hot encoding | String |
| N29 | Li | Node feature | One hot encoding | String |
| N30 | Ge | Node feature | One hot encoding | String |
| N31 | Cu | Node feature | One hot encoding | String |
| N32 | Au | Node feature | One hot encoding | String |
| N33 | Ni | Node feature | One hot encoding | String |
| N34 | Cd | Node feature | One hot encoding | String |
| N35 | In | Node feature | One hot encoding | String |
| N36 | Mn | Node feature | One hot encoding | String |
| N37 | Zr | Node feature | One hot encoding | String |

| N38 | Cr | Node feature | One hot encoding | String |
|---|---|---|---|---|
| N39 | Pt | Node feature | One hot encoding | String |
| N40 | Hg | Node feature | One hot encoding | String |
| N4 | Pb | Node feature | One hot encoding | String |
| N42 | Unknown element | Node feature | One hot encoding | String |
| N43 | n_heavy_neighbors_0 | Node feature | One hot encoding | String |
| N44 | n_heavy_neighbors_1 | Node feature | One hot encoding | String |
| N45 | n_heavy_neighbors_2 | Node feature | One hot encoding | String |
| N46 | n_heavy_neighbors_3 | Node feature | One hot encoding | String |
| N47 | n_heavy_neighbors_4 | Node feature | One hot encoding | String |
| N48 | n_heavy_neighbors_MoreThanFour | Node feature | One hot encoding | String |
| N49 | formal_charge_-3 | Node feature | One hot encoding | String |
| N50 | formal_charge_-2 | Node feature | One hot encoding | String |
| N51 | formal_charge_-1 | Node feature | One hot encoding | String |
| N52 | formal_charge_0 | Node feature | One hot encoding | String |
| N53 | formal_charge_1 | Node feature | One hot encoding | String |
| N54 | formal_charge_2 | Node feature | One hot encoding | String |
| N55 | formal_charge_3 | Node feature | One hot encoding | String |
| N56 | formal_charge_Extreme | Node feature | One hot encoding | String |
| N57 | S | Node feature | One hot encoding | String |
| N58 | SP | Node feature | One hot encoding | String |
| N59 | SP2 | Node feature | One hot encoding | String |
| N60 | SP3 | Node feature | One hot encoding | String |
| N61 | SP3D | Node feature | One hot encoding | String |
| N62 | SP3D2 | Node feature | One hot encoding | String |
| N63 | Other hybridization | Node feature | One hot encoding | String |
| N64 | is_in_a_ring | Node feature | Numerical | Int |
| N65 | is_aromatic | Node feature | Numerical | Int |
| N66 | atomic_mass_scaled | Node feature | Numerical | Float |
| N67 | vdw_radius_scaled | Node feature | Numerical | Float |
| N68 | covalent_radius_scaled | Node feature | Numerical | Float |
| N69 | CHI_UNSPECIFIED | Node feature | One hot encoding | String |
| N70 | CHI_TETRAHEDRAL_CW | Node feature | One hot encoding | String |
| N71 | CHI_TETRAHEDRAL_CCW | Node feature | One hot encoding | String |
| N72 | CHI_OTHER | Node feature | One hot encoding | String |
| N73 | n_hydrogens_0 | Node feature | One hot encoding | String |
| N74 | n_hydrogens_1 | Node feature | One hot encoding | String |
| N75 | n_hydrogens_2 | Node feature | One hot encoding | String |
| N76 | n_hydrogens_3 | Node feature | One hot encoding | String |

| N77 | n_hydrogens_4 | Node feature | One hot encoding | String |
|---|---|---|---|---|
| N78 | n_hydrogens_MoreThanFour | Node feature | One hot encoding | String |

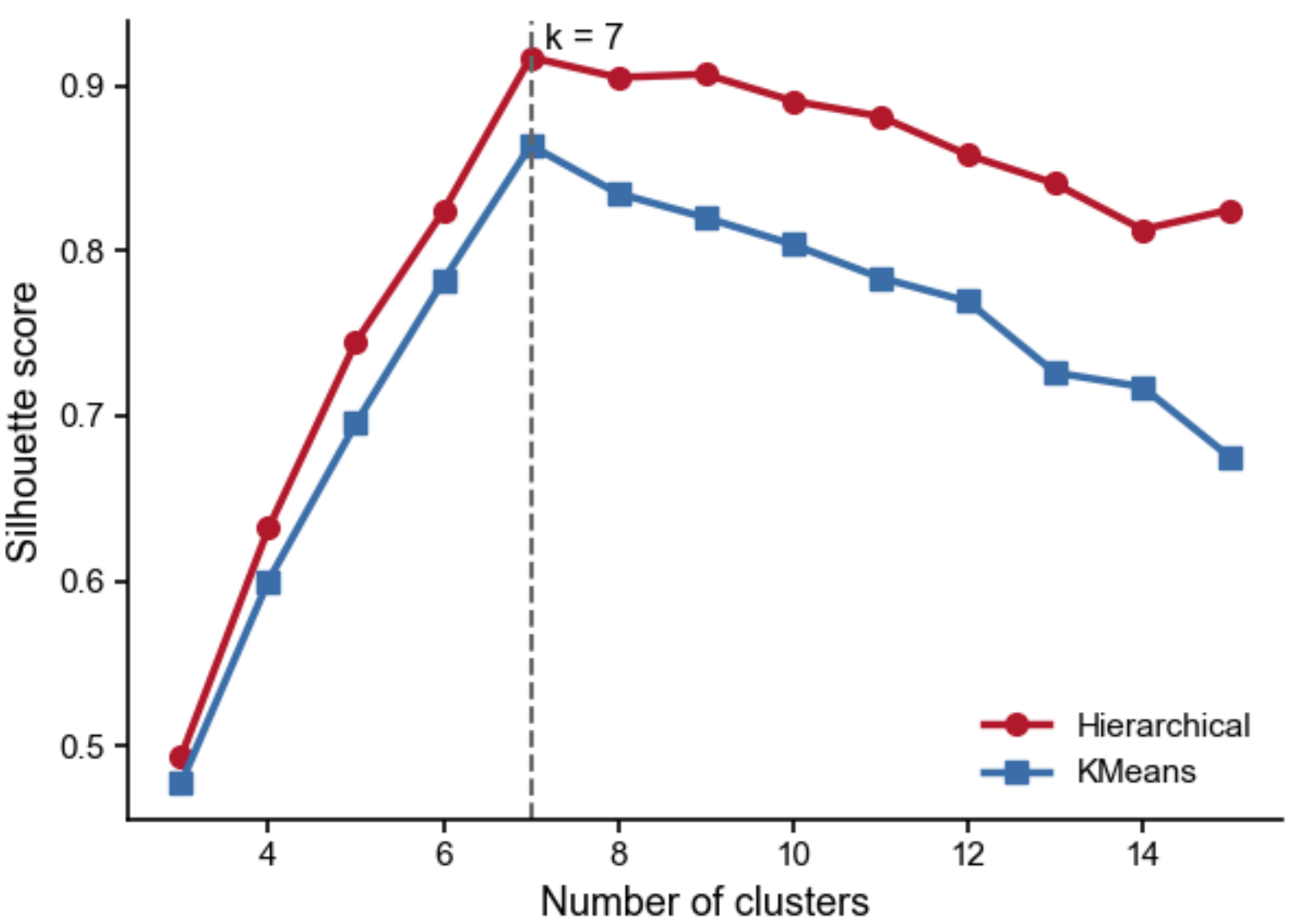

**Figure S2 | Selection of the optimal number of molecular prototype.** Average silhouette scores of hierarchical clustering and K-means[3] clustering evaluated for cluster numbers ranging from 3 to 15. The optimal number of molecular prototypes was determined using the silhouette score, which measures the compactness and separation of molecular clusters. Hierarchical clustering achieved the highest silhouette score at k = 7, while K-means clustering showed a consistent optimum in the same region. Therefore, seven molecular prototypes were selected in a data-driven manner and used throughout this study for unsupervised prototype discovery and subsequent prototype-guided learning.

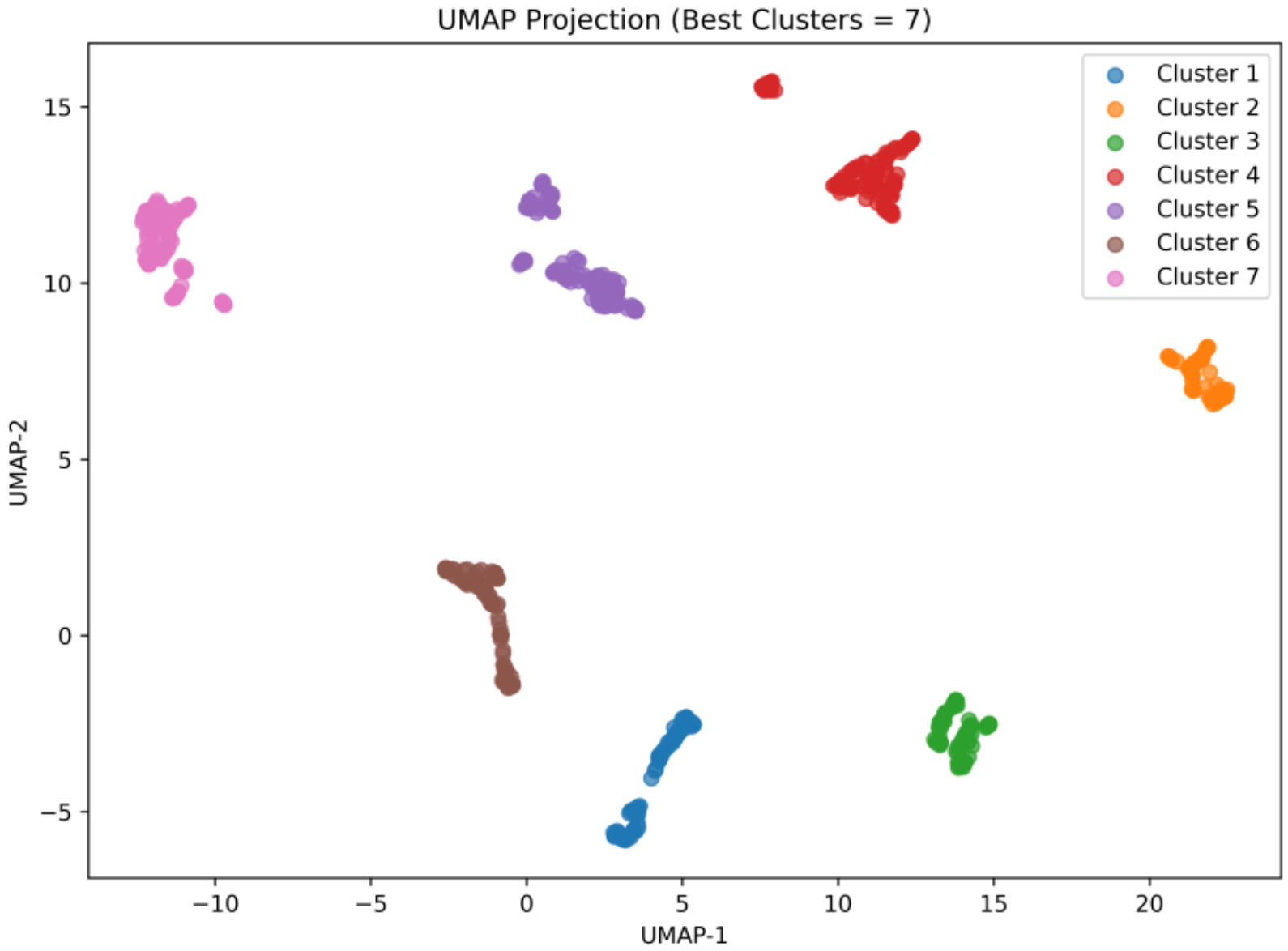

**Figure S3 | UMAP[4] visualization of the reported boron electrolyte additive molecules learned by the UCL model.** UMAP projection of the learned graph-level embeddings of reported boron-containing electrolyte additives. Each point represents one reported additive molecule, and colors denote the prototype assignments obtained by

hierarchical clustering in the learned representation space. The seven well-separated groups indicate that the UCL model organizes reported additives into distinct molecular prototype regions, consistent with the optimal cluster number determined by silhouette-score analysis in Supplementary Figure S2. The spatial separation of the clusters supports the presence of structurally and functionally distinguishable additive prototypes encoded in the learned embedding space.

**Table S3 | Reported additive molecules in each prototypes.**

| Molecular ID | Molecular name | Molecules | Molecular weight | Synthetic accessibility Score[5] | Prototype label | References |
|---|---|---|---|---|---|---|
| 6 | TBDB | | 225.890 | 4.273 | 1 | 6 |
| 10 | ITD | | 186.060 | 3.616 | 1 | 7 |
| 12 | SOFPB | | 352.331 | 2.987 | 1 | 8 |
| 14 | DPD-CN-OCH3 | | 245.087 | 3.159 | 1 | 9 |
| 19 | C-LiMCFB | | 317.089 | 4.211 | 1 | 10 |
| 24 | NHCABT | | 167.971 | 3.293 | 1 | 11 |
| 32 | DPD | | 190.051 | 2.893 | 1 | 9 |
| 43 | BF3-THF | | 139.913 | 5.532 | 1 | 12 |
| 59 | DiOB-Py | | 205.066 | 2.853 | 1 | 13 |
| 66 | BEG-1 | | 412.185 | 4.069 | 1 | 14 |
| 83 | DPD-CN | | 215.061 | 3.159 | 1 | 9 |
| 92 | DPD-F | | 208.041 | 2.98 | 1 | 9 |
| 95 | BDB | | 253.944 | 3.428 | 1 | 15 |
| 98 | DPD-CHO | | 218.061 | 3.128 | 1 | 9 |
| 104 | DPD-COOCH2CH3 | | 262.114 | 2.852 | 1 | 9 |
| 116 | DiOB-An | | 275.201 | 2.699 | 1 | 13 |
| 3 | PFPBO | | 265.886 | 3.559 | 2 | 16 |
| 34 | DCLBSB | | 427.745 | 4.164 | 2 | 17 |
| 45 | TFPBO | | 559.977 | 2.965 | 2 | 18 |
| 51 | LiBBrSB |  | 462.792 | 4.368 | 2 | [19] |
| 54 | PFPTFBB |  | 357.924 | 3.393 | 2 | [20] |
| 56 | TPFPB |  | 511.980 | 2.800 | 2 | [21] |
| 75 | TPF |  | 511.980 | 2.800 | 2 | [22] |
| 77 | TFPB |  | 511.980 | 2.800 | 2 | [18] |
| 94 | TCLBSB |  | 496.635 | 4.388 | 2 | [17] |
| 109 | 4FLBDOB |  | 285.827 | 4.647 | 2 | [23] |
| 115 | LiBClSB |  | 373.89 | 4.260 | 2 | [19] |
| 122 | LiBCl2SB |  | 442.78 | 4.355 | 2 | [19] |
| 4 | LiPVAOB |  | 221.931 | 5.608 | 3 | [24] |
| 30 | ADM |  | 170.961 | 3.811 | 3 | [25] |
| 36 | LiODFB | | 143.767 | 4.662 | 3 | 26 |
| 37 | LiPAAOB | | 278.959 | 4.432 | 3 | 27 |
| 52 | LBDCB | | 265.899 | 5.246 | 3 | 28 |
| 61 | LDFOB | | 143.767 | 4.662 | 3 | 29 |
| 67 | LBCB | | 297.853 | 5.653 | 3 | 28 |
| 69 | NaDFOB | | 159.816 | 4.729 | 3 | 30 |
| 73 | LiBOB | | 193.789 | 4.913 | 3 | 19 |
| 101 | LiMOB | | 221.843 | 4.983 | 3 | 31 |
| 102 | LiBMAB | | 301.973 | 4.961 | 3 | 32 |
| 105 | LOCB | | 245.821 | 5.846 | 3 | 28 |

| 110 | LiBF4 | | 193.789 | 4.913 | 3 | [29] |
|---|---|---|---|---|---|---|
| 112 | LiDFOB | | 143.767 | 4.662 | 3 | [33] |
| 5 | (C6H3F)O2B(C8H3 F6) | | 349.998 | 3.089 | 4 | [34] |
| 8 | LBBPB | | 386.141 | 3.111 | 4 | [35] |
| 9 | TPhBX | | 311.751 | 2.739 | 4 | [36] |
| 15 | (C6H3F)O2B(C6H3 F2) | | 249.984 | 2.911 | 4 | [34] |
| 18 | LBSB | | 289.965 | 3.837 | 4 | [17] |
| 28 | LBNB | | 334.065 | 3.487 | 4 | [35] |
| 50 | (C6H3F)O2B(C7H4 F3) | | 282.001 | 2.880 | 4 | [34] |
| 55 | (C6F4)O2B(C6F5) | | 234.010 | 2.888 | 4 | [34] |
| 60 | (C6F4)O2B(C8H3F 6) | | 351.006 | 3.059 | 4 | [34] |
| 70 | LBDOB | | 233.945 | 3.733 | 4 | [37] |
| 76 | LBBB | | 233.945 | 3.733 | 4 | [35] |
| 82 | (C6F4)O2B(C7H3F6) | | 283.009 | 2.906 | 4 | [34] |
| 84 | LiBPh4 | | 326.177 | 2.093 | 4 | [38] |
| 85 | 3-MLBSB | | 318.019 | 4.029 | 4 | [17] |
| 88 | (C6F4)O2B(C6H4F) | | 233.002 | 2.742 | 4 | [34] |
| 89 | Li[BScB] | | 289.965 | 3.837 | 4 | [39] |
| 97 | B(OPh)3 | | 290.127 | 2.012 | 4 | [40] |
| 107 | LiBSB | | 289.965 | 3.837 | 4 | [19] |
| 113 | (C6F4)O2B(C6H3F2) | | 250.992 | 2.873 | 4 | [34] |
| 7 | TRIFM | | 214.905 | 2.601 | 5 | [41] |
| 11 | LBO | | 49.751 | 5.870 | 5 | 42 |
| 21 | Li2TB | | 169.123 | 6.757 | 5 | 43 |
| 33 | PRZ | | 215.702 | 4.900 | 5 | 44 |
| 35 | H3BO3 | | 61.833 | 3.432 | 5 | 45 |
| 38 | FBA | | 189.929 | 2.415 | 5 | 46 |
| 41 | Li3BO3 | | 65.750 | 6.320 | 5 | 47 |
| 44 | B2O3 | | 263811 | 5.773 | 5 | 48 |
| 57 | LiBO2 | | 65.750 | 6.320 | 5 | 49 |
| 58 | TEB | | 65.750 | 6.320 | 5 | 50 |
| 65 | TMSPB | | 194.115 | 2.914 | 5 | 51 |
| 71 | MDMB | | 192.067 | 2.911 | 5 | 52 |
| 74 | 3F-PBF | | 164.898 | 3.899 | 5 | [41] |
| 78 | LUTID | | 174.962 | 3.700 | 5 | [41] |
| 86 | PBF | | 146.908 | 2.415 | 5 | [41] |
| 91 | TRIZIN | | 284.496 | 5.317 | 5 | [53] |
| 96 | K2B4O5(OH)4 | | 704.412 | 6.036 | 5 | [54] |
| 103 | VPBF | | 172.946 | 2.874 | 5 | [41] |
| 106 | DEBA | | 99.926 | 4.079 | 5 | [55] |
| 111 | PCN | | 171.918 | 2.711 | 5 | [41] |
| 121 | LiB5AB | | 385.906 | 3.065 | 5 | [56] |
| 124 | 2F-PBF | | 164.898 | 2.703 | 5 | [41] |
| 13 | (C3HF6O)2B(C6H5) | | 421.974 | 3.417 | 6 | [34] |
| 17 | THFBuBO | | 607.947 | 3.432 | 6 | 57 |
| 25 | (C3HF6O)2B(C6F5) | | 440.972 | 3.547 | 6 | 34 |
| 31 | (C3HF6O)2B(C6H3F2) | | 457.954 | 3.594 | 6 | 34 |
| 46 | TFEB | | 307.905 | 3.422 | 6 | 43 |
| 63 | B(HFIP)3 | | 511.896 | 3.443 | 6 | 58 |
| 68 | Na[B(hfip)4] | | 701.914 | 3.743 | 6 | 59 |
| 90 | THFB | | 511.896 | 3.687 | 6 | 57 |
| 93 | TTFEB | | 307.905 | 3.422 | 6 | 60 |
| 99 | TMF | | 613.863 | 3.644 | 6 | 22 |
| 108 | TTFB | | 307.905 | 3.422 | 6 | 61 |
| 119 | ABA#1 | | 511.896 | 3.443 | 6 | 62 |
| 123 | THFPB | | 511.896 | 3.443 | 6 | 63 |
| 125 | THB | | 511.896 | 3.443 | 6 | 64 |
| 0 | TMSB | | 278.382 | 3.985 | 7 | 65 |
| 1 | TCEB | | 221.025 | 3.847 | 7 | 66 |
| 2 | TMBx | | 173.535 | 5.202 | 7 | 67 |
| 16 | TiPBx | | 257.697 | 4.618 | 7 | 67 |
| 20 | L-LiMCFB | | 326.046 | 3.152 | 7 | 10 |
| 22 | TME | | 188.076 | 3.621 | 7 | 22 |
| 26 | HBC | | 1455.489 | 4.462 | 7 | 68 |
| 27 | TnPBx | | 257.697 | 4.219 | 7 | 67 |
| 29 | LBC | | 1059.012 | 3.722 | 7 | 68 |

| 40 | LiT4PAB | | 414.189 | 3.856 | 7 | [69] |
|---|---|---|---|---|---|---|
| 47 | BF3·DE | | 141.929 | 2.963 | 7 | [70] |
| 48 | BPC | | 1060.984 | 3.801 | 7 | [71] |
| 64 | TPB | | 188.076 | 3.388 | 7 | [72] |
| 72 | TAB | | 342.197 | 3.497 | 7 | [73] |
| 79 | TMB | | 125.538 | 5.382 | 7 | [74] |
| 80 | B-HEMA | | 398.217 | 3.249 | 7 | [68] |
| 100 | PBA | | 137.931 | 2.255 | 7 | [75] |
| 114 | TBB | | 230.157 | 2.858 | 7 | [76] |
| 117 | TEBx | | 215.616 | 4.804 | 7 | [67] |

The missing compound ID was filtered out as an outlier during the clustering process.

**Table S4 | Shared motif summary table**

| Legend in Fig. 3e | Motif name | Motif structure | Appears in the prototype |
|---|---|---|---|
| P1-SM | C-B(OR)2 motif | | P1, P2, P3, P4, P5, P6 |
| P2-SM | Pentafluorophenyl group | | P2 |
| P3-SM | Boron oxalate | | P2, P3 |
| P4-SM | Catechol boronate | | P2, P3, P4 |
| P5-SM | BF3-like motif | | P1, P5, P7 |
| P6-SM | Fluorinated alkoxy | | P6 |
| P7-SM | Tri-alkoxy borate | | P1, P2, P3, P4, P5, P6, P7 |
| 8-SM | Tetra-alkoxy borate | | P2, P3, P4, P6, P7 |

The shared molecular framework was identified using the SMART[77] method.

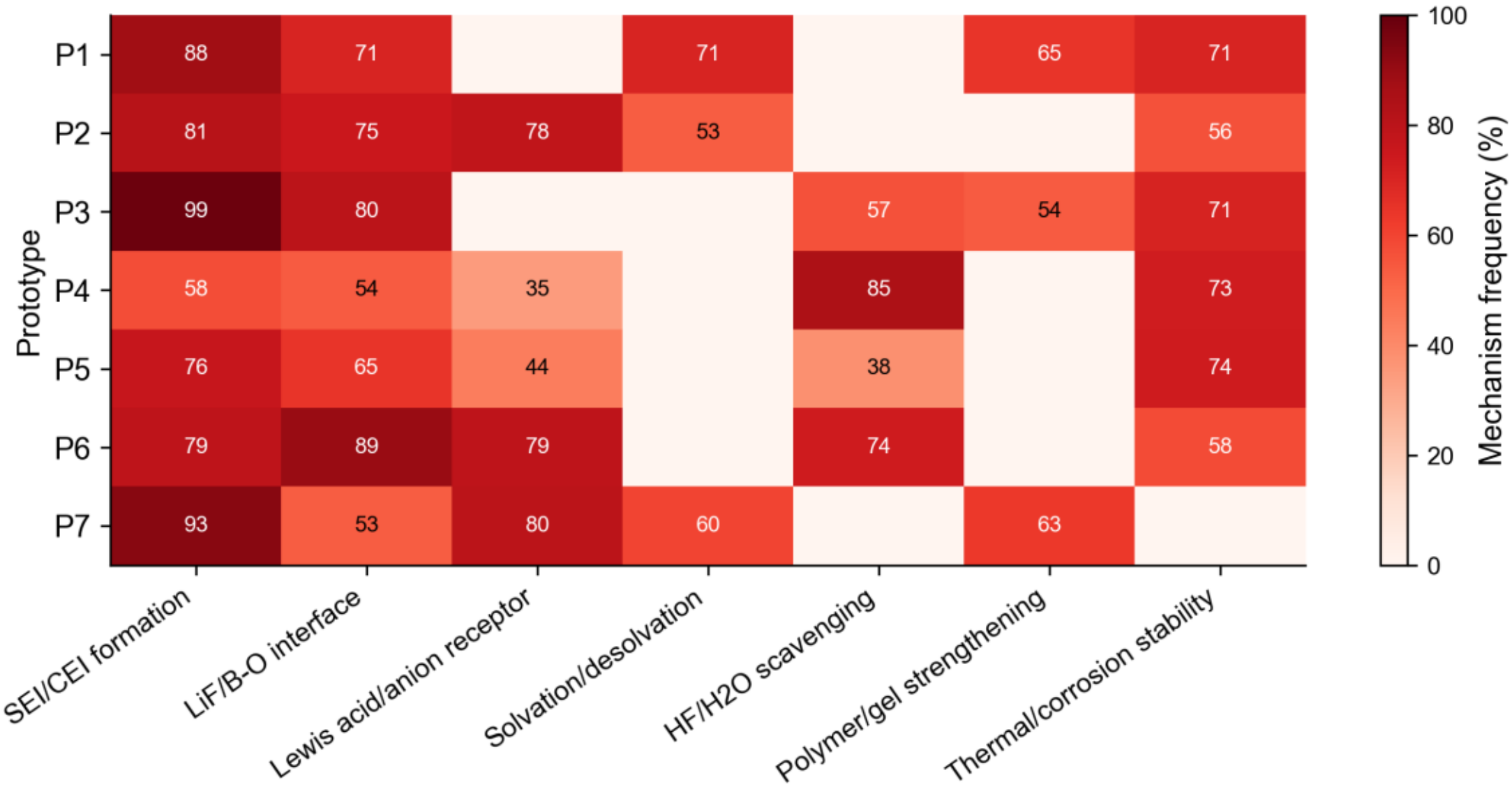

**Figure S4 | Mechanism annotation statistics of the seven learned molecular prototypes.** Quantitative mechanism annotation supporting the functional profiles in Fig. 3c. Heatmap showing the normalized frequency of seven electrolyte-regulation mechanisms across the seven learned molecular prototypes. Frequencies were calculated by dividing the number of mechanism-related records by the total number of annotated records in each prototype.

**Table S5 | Prototypes summary table**

| Prototype | Number of molecules | Representative molecule | Dominant motifs |
|---|---|---|---|
| P1 | 89 | DiOB-Py | C-B(OR)2 motif |
| P2 | 64 | TPFPB | pentafluorophenyl group |
| P3 | 78 | LiDFOB | tetra-alkoxy borate |
| P4 | 117 | LBBB | catechol boronate |
| P5 | 131 | PCN | BF3-like motif |
| P6 | 92 | TTFEB | fluorinated alkoxy |
| P7 | 109 | TBB | tri-alkoxy borate |

The population is calculated after the graph augmentation.

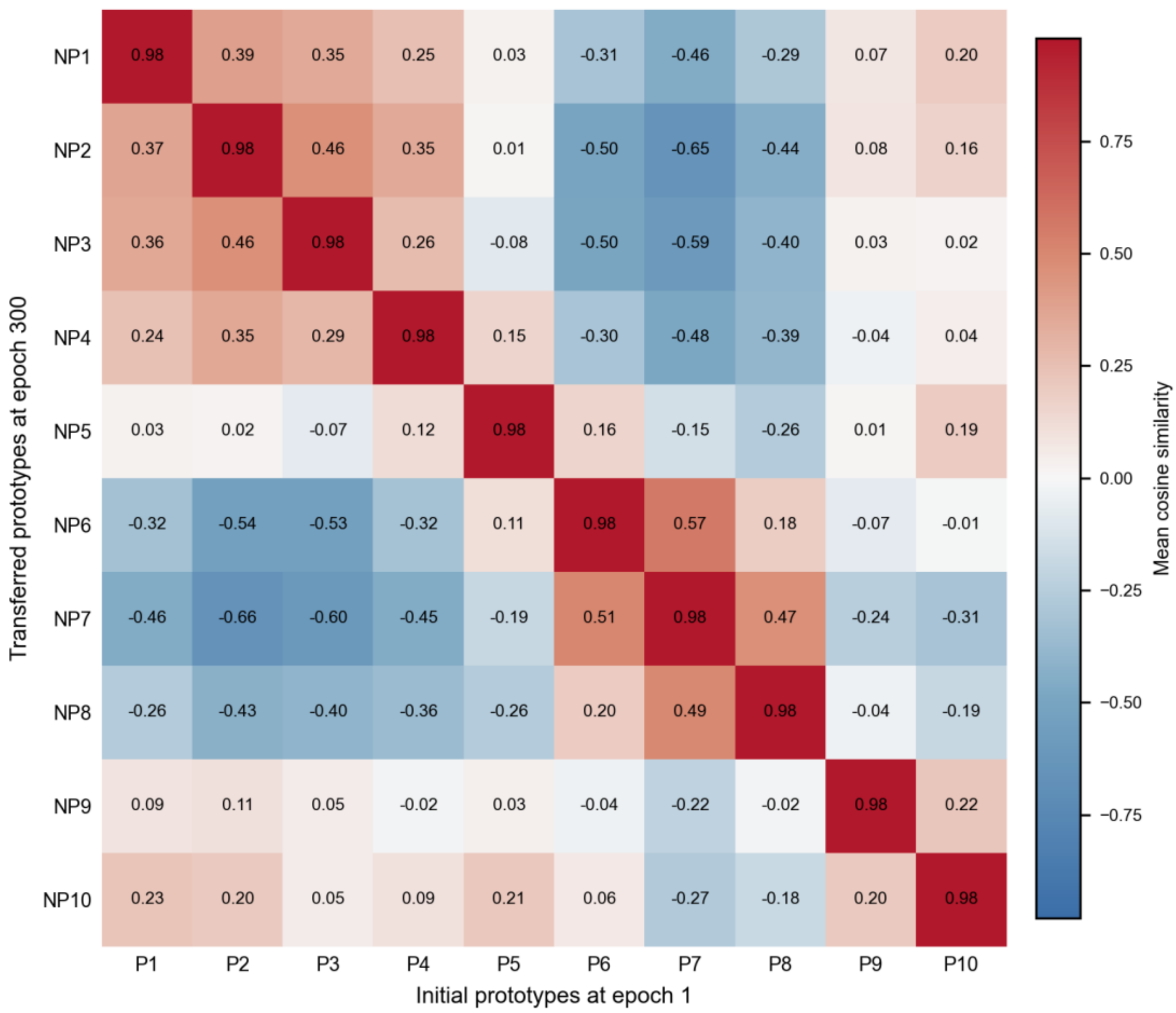

**Figure S5 | Prototype correspondence between initial and transferred prototype centroids.** Heatmap showing the mean cosine similarity between the initial prototype centroids at epoch 1 and the transferred prototype centroids after PCL training at epoch 300. Rows denote transferred prototypes, and columns denote initial prototypes. The strong diagonal similarity indicates that most transferred prototypes preserve clear correspondence with their initial prototype identities, whereas structured off-diagonal similarities reflect partial reorganization and information sharing among related prototype regions during adaptation. These results support that PCL maintains prototype-level continuity while allowing the prototype organization to adjust to the unlabeled candidate chemical space.

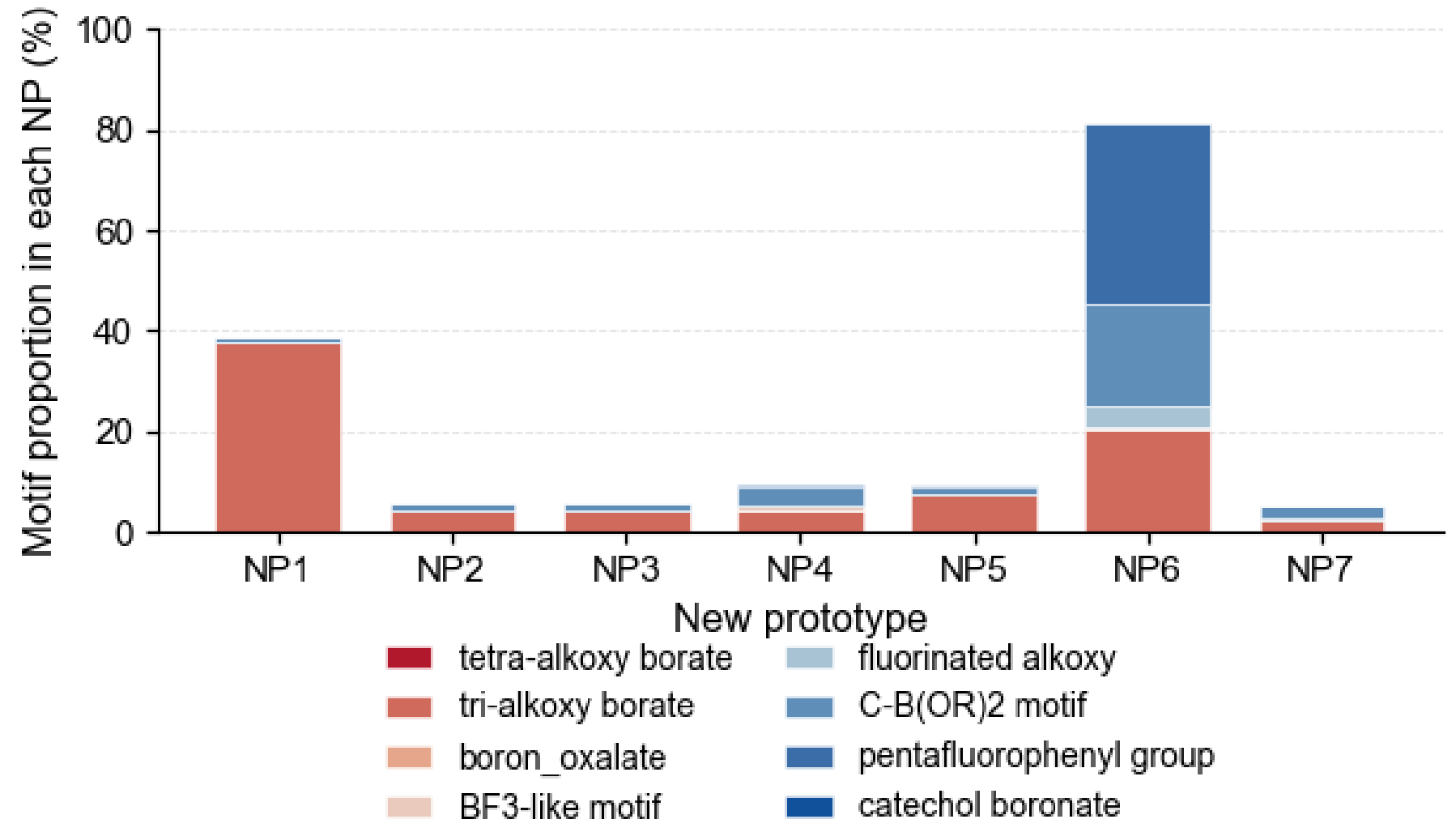

**Figure S6 | Structural motif distribution across transferred prototypes.** Stacked bar plot showing the distribution of representative boron-containing structural motifs across transferred prototypes after PCL adaptation. Motif proportions were calculated from molecules assigned to each new prototype using predefined SMiles ARbitrary Target Specification (SMARTS) patterns, including tetra-alkoxy borate, tri-alkoxy borate, boron oxalate, BF3-like motifs, fluorinated alkoxy groups, C-B(OR)2 motifs, pentafluorophenyl groups and catechol boronate. The enrichment of different motifs in distinct transferred prototypes indicates that PCL preserves chemically meaningful structural information during prototype adaptation rather than forming arbitrary clusters in the candidate space. Because a molecule may contain more than one motif, stacked heights represent cumulative motif occurrence rather than mutually exclusive molecular classes.

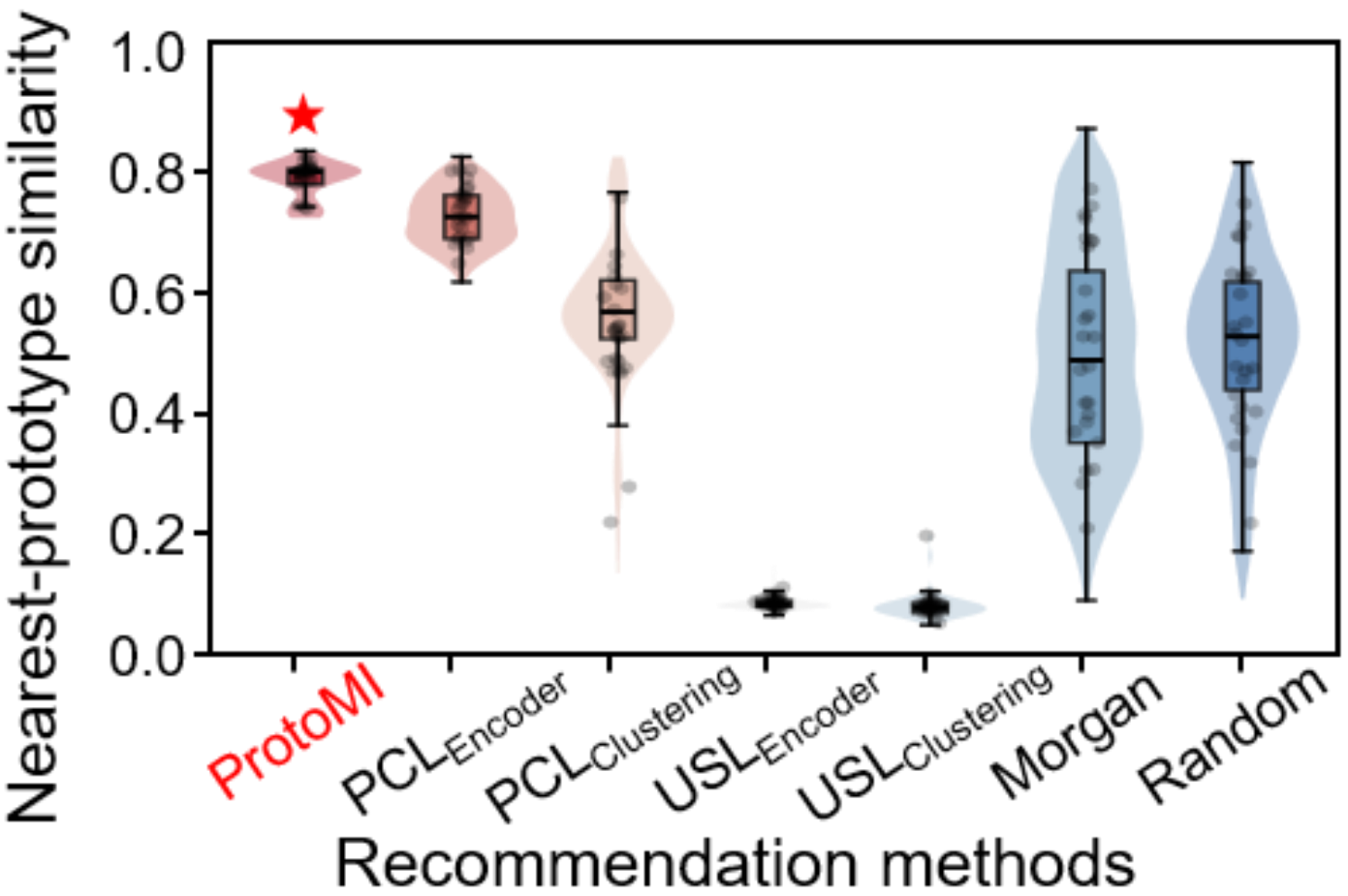

**Figure S7 | Distribution of prototype similarity across recommendation strategies.** Violin plots showing the distribution of molecule-level prototype similarity for molecules recommended by different strategies after translation strategy. Each point represents one recommended molecule, and box plots indicate the median and interquartile range. ProtoMI shows the highest overall prototype similarity among the compared strategies, indicating that prototype-guided adaptation enriches molecules that remain close to the learned additive prototype regions. The red star highlights ProtoMI achieves the highest prototype similarity among all recommendation methods.

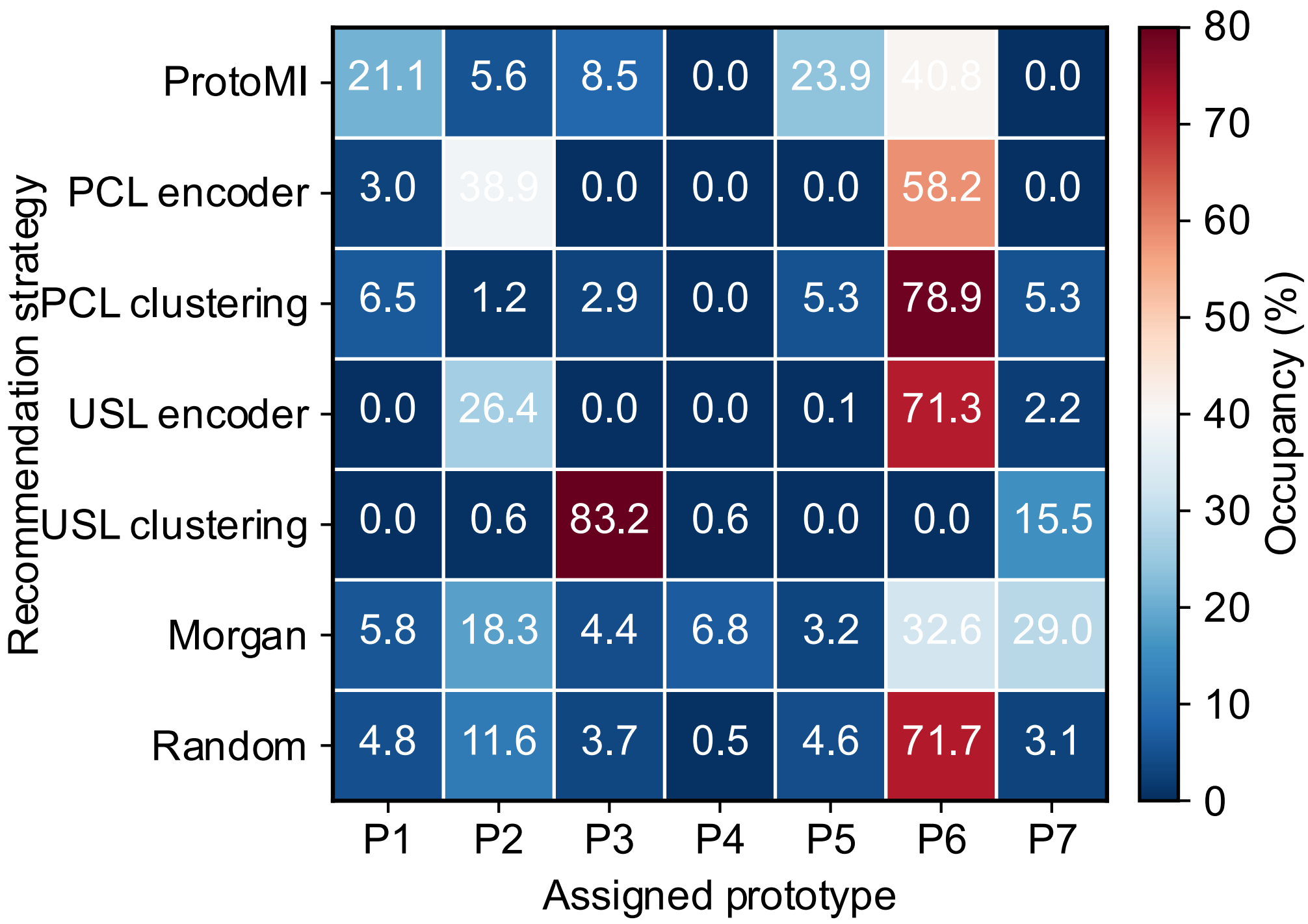

**Figure S8 | Prototype occupancy of molecules recommended by different strategies.** Heatmap showing the occupancy of recommended molecules across the seven learned molecular prototypes after translation strategy. Rows correspond to different recommendation strategies, and columns correspond to assigned prototypes P1–P7. Values indicate the percentage of molecules assigned to each prototype within a given strategy, with each row normalized to 100%. ProtoMI distributes recommended molecules across multiple prototype regions while retaining high prototype similarity, suggesting that the model avoids collapse into a single prototype and preserves chemically diverse recommendation coverage.

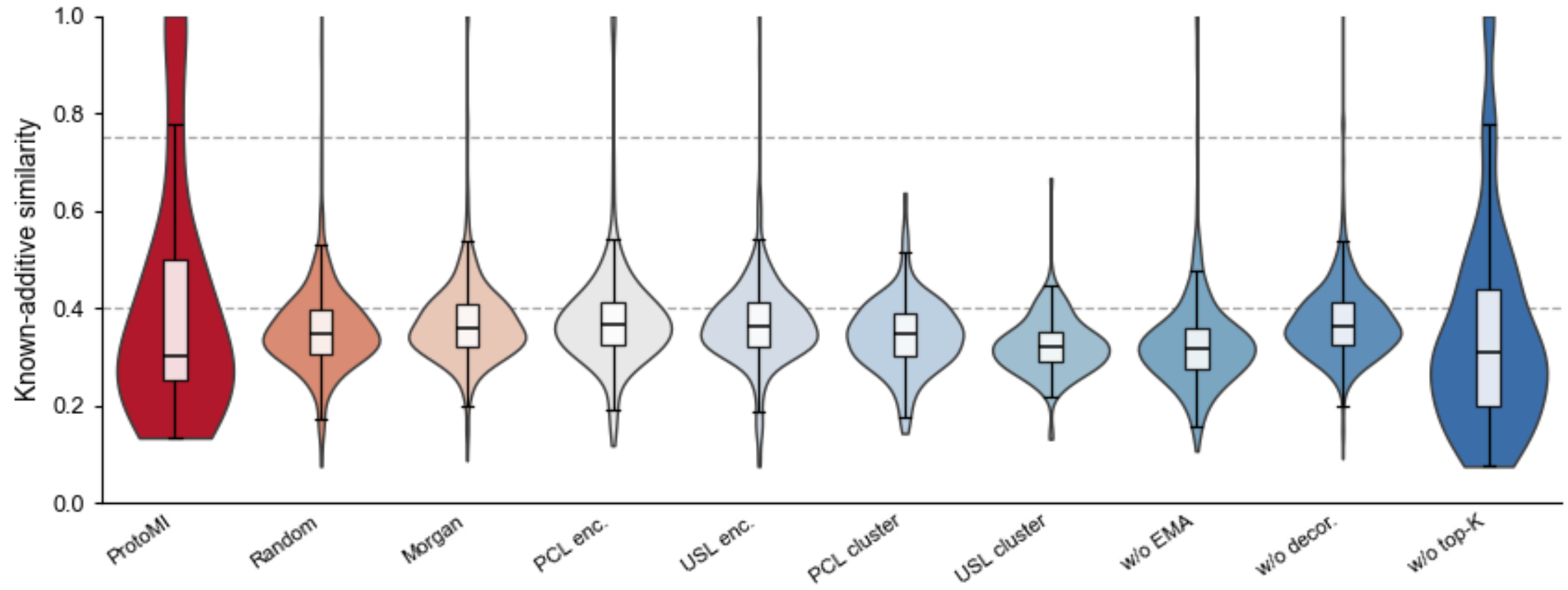

**Figure S9 | Similarity of recommended molecules to reported additives.** Distribution of the maximum structural similarity between recommended molecules and reported boron-containing electrolyte additives across different recommendation strategies. Each violin plot shows the molecule-level known-additive similarity after post-screening, with box plots indicating the median and interquartile range. ProtoMI exhibits a broad similarity distribution, including both candidates close to reported additives and candidates with lower similarity, suggesting that the model transfers prototype-level chemical knowledge without simply reproducing close analogues of known molecules.

**Table S6 | Extra recommendation statistics**

| Method | Entropy-weighted prototype similarity | Novelty score | High similarity ratio >0.75 | Moderate similarity ratio 0.40-0.75 | Low similarity ratio <0.40 | Post-screening survival |
|---|---|---|---|---|---|---|
| Random selection | 27.637 | 0.6435 ± 0.0910 | 0.62% | 24.42% | 74.96% | 2.10% |
| Morgan fingerprints | 41.600 | 0.6304 ± 0.0875 | 0.75% | 29.10% | 70.15% | 5.18% |
| UCL encoder only | 2.985 | 0.6323 ± 0.0889 | 0.43% | 30.70% | 68.87% | 5.26% |
| PCL encoder only | 29.466 | 0.6206 ± 0.1127 | 1.73% | 29.21% | 69.06% | 1.31% |
| UCL encoder clustering | 2.079 | 0.6762 ± 0.0607 | 0.00% | 10.32% | 89.68% | 0.50% |
| PCL encoder clustering | 24.364 | 0.6551 ± 0.0744 | 0.00% | 22.29% | 77.71% | 1.10% |
| ProtoMI | 57.121 | 0.5975 ± 0.2446 | 12.68% | 25.35% | 61.97% | 0.23% |
| w/o EMA | 9.456 | 0.6722 ± 0.1029 | 1.22% | 12.91% | 85.86% | 1.86% |
| w/o decorrelation | 0.299 | 0.6270 ± 0.0817 | 0.75% | 30.21% | 69.04% | 5.65% |
| w/o top-k | 0.000 | 0.6558 ± 0.2060 | 7.07% | 23.91% | 69.02% | 0.96% |

**Table S7 | Full temporal validation**

| Cutoff year | Recall | Precision | F1 score | Enrichment factor | Reduction factor | Hit number |
|---|---|---|---|---|---|---|
| 2017 | 0.10638 | 0.00274 | 0.00535 | 10.52583 | 98.94282 | 5/47 |
| 2019 | 0.11627 | 0.00219 | 0.00430 | 9.18278 | 78.97191 | 5/43 |
| 2021 | 0.08823 | 0.00862 | 0.01570 | 45.63311 | 517.17528 | 3/34 |
| 2023 | 0.20000 | 0.00484 | 0.00946 | 43.63078 | 218.15393 | 4/20 |

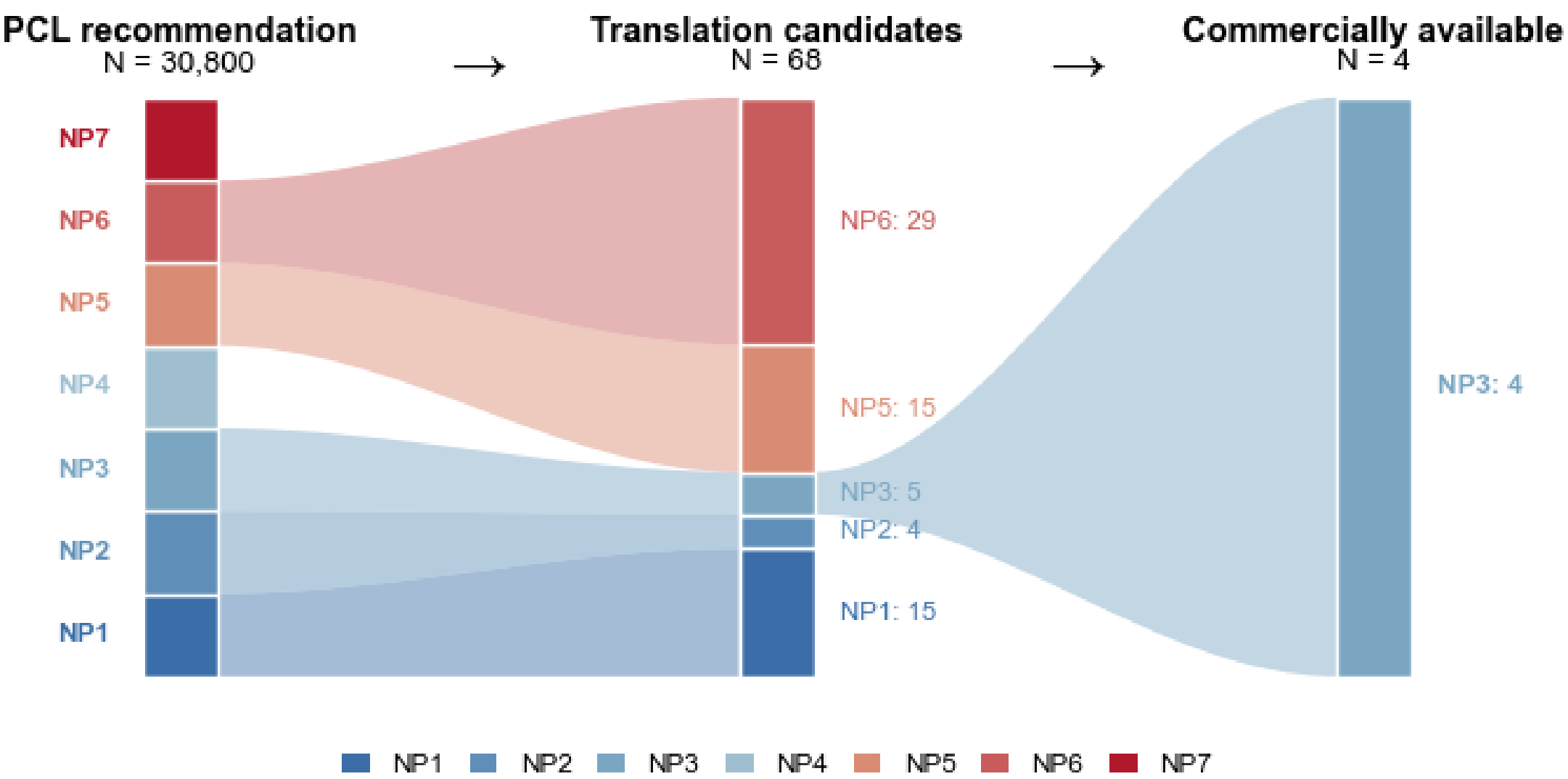

**Figure S10 | Prototype survival during translation from model recommendations to experimentally actionable candidates.** Sankey diagram showing the retention and redistribution of new prototype assignments during the translation strategy from PCL-recommended molecules to experimentally actionable candidates. The initial PCL recommendation pool contained 30,800 molecules distributed across seven new prototypes. Sequential filtering based on molecular validity, molecular weight, synthetic accessibility, active hydrogen exclusion, CAS availability and practical cost considerations reduced the pool to 68 translation candidates. Further prioritization according to commercial availability and experimental feasibility yielded four candidates for electrochemical validation, all assigned to NP3. Flow widths are proportional to the number of molecules retained from each prototype at each filtering stage.

**Table S8 | 68 Candidates after translation strategy**

| Index | Compound ID | Formula | Molecular weight | SA Score | Prototype label | CAS number | Reported or not |
|---|---|---|---|---|---|---|---|
| 1 | 11090867 | | 409.053 | 3.177 | 1 | 119618-67-6 | No |
| 2 | 15877487 | | 180.095 | 2.739 | 1 | 253280-01-2 | No |
| 3 | 57041839 | | 292.148 | 3.585 | 1 | 62043-02-1 | No |
| 4 | 71390453 | | 292.148 | 3.624 | 1 | 62043-03-2 | No |
| 5 | 78020 | | 278.136 | 3.984 | 1 | 4325-85-3 | Yes |
| 6 | 425307 | | 188.049 | 3.456 | 1 | 4887-24-5 | No |
| 7 | 12418354 | | 266.132 | 3.356 | 1 | 7560-51-2 | No |
| 8 | 22338812 | | 266.092 | 3.121 | 1 | 55923-08-5 | No |
| 9 | 56973279 | | 344.957 | 3.429 | 1 | 1356165-73-5 | No |
| 10 | 517955 | | 69.940 | 3.831 | 1 | 1113-22-0 | No |
| 11 | 138247 | | 71.920 | 3.989 | 1 | 4443-43-0 | No |
| 12 | 13301362 | | 222.050 | 3.060 | 1 | 94839-08-4 | No |
| 13 | 13789232 | | 239.500 | 3.942 | 1 | 107134-81-6 | No |
| 14 | 68979 | | 55.920 | 3.834 | 1 | 593-90-8 | No |
| 15 | 25068199 | | 405.100 | 3.195 | 1 | 951163-66-9 | No |
| 16 | 66709077 | | 182.122 | 3.664 | 2 | 1392108-88-1 | No |
| 17 | 15158954 | | 148.020 | 3.571 | 2 | 139298-13-8 | No |
| 18 | 11522786 | | 331.000 | 2.822 | 2 | 872044-92-3 | No |
| 19 | 22663171 | | 193.120 | 3.577 | 2 | 141938-41-2 | No |
| 20 | 25112575 | | 556.257 | 2.677 | 3 | 624744-67-8 | No |
| 21 | 58050304 | | 430.210 | 2.540 | 3 | 1149804-35-2 | No |
| 22 | 58489416 | | 430.210 | 2.524 | 3 | 1115639-92-3 | No |
| 23 | 23115783 | | 502.500 | 2.529 | 3 | 281668-51-7 | No |
| 24 | 58769449 | | 354.300 | 2.472 | 3 | 890042-13-4 | No |
| 25 | 21425 | | 818.862 | 2.456 | 5 | 5337-41-7 | No |
| 26 | 125587 | | 224.200 | 3.947 | 5 | 32327-52-9 | No |
| 27 | 13680338 | | 238.300 | 3.883 | 5 | 16413-17-5 | No |
| 28 | 71384846 | | 350.500 | 3.582 | 5 | 62594-01-8 | No |
| 29 | 71384845 | | 350.500 | 3.983 | 5 | 62594-02-9 | No |
| 30 | 71334465 | | 560.900 | 2.331 | 5 | 106787-52-4 | No |
| 31 | 71386315 | | 460.700 | 2.844 | 5 | 62444-57-9 | No |
| 32 | 71352225 | | 476.700 | 2.070 | 5 | 1188-97-2 | No |
| 33 | 352104 | | 518.799 | 2.065 | 5 | 14245-38-6 | No |
| 34 | 71346374 | | 211.180 | 3.403 | 5 | 151293-65-1 | No |
| 35 | 17177 | | 398.500 | 3.931 | 5 | 2467-13-2 | No |
| 36 | 21424 | | 314.300 | 3.996 | 5 | 5337-37-1 | No |
| 37 | 17178 | | 566.800 | 2.247 | 5 | 2467-15-4 | No |
| 38 | 219443 | | 440.600 | 3.010 | 5 | 3088-77-5 | No |
| 39 | 108841 | | 608.900 | 2.480 | 5 | 59802-06-1 | No |
| 40 | 15605783 | | 296.100 | 3.454 | 6 | 267006-38-2 | No |
| 41 | 18373965 | | 240.038 | 3.337 | 6 | 848609-02-9 | No |
| 42 | 582056 | | 511.985 | 2.800 | 6 | 1109-15-5 | Yes |
| 43 | 11727684 | | 344.993 | 2.819 | 6 | 165612-94-2 | No |
| 44 | 75360847 | | 458.017 | 3.449 | 6 | 2003-04-5 | Yes |
| 45 | 10959856 | | 380.380 | 3.032 | 6 | 2720-03-8 | No |
| 46 | 11115716 | | 359.960 | 2.958 | 6 | 170151-48-1 | No |
| 47 | 12626668 | | 248.770 | 3.405 | 6 | 830-48-8 | No |
| 48 | 11050988 | | 770.100 | 3.300 | 6 | 190282-03-2 | Yes |
| 49 | 15520532 | | 207.940 | 3.206 | 6 | 363596-54-7 | No |
| 50 | 10876447 | | 770.100 | 3.354 | 6 | 204930-04-1 | Yes |
| 51 | 71428744 | | 494.000 | 3.182 | 6 | 916442-23-4 | Yes |
| 52 | 71420187 | | 598.200 | 3.056 | 6 | 820972-26-7 | No |
| 53 | 71447200 | | 367.850 | 3.257 | 6 | 329009-01-0 | No |
| 54 | 71350376 | | 448.100 | 3.152 | 6 | 165612-90-8 | No |
| 55 | 71428745 | | 854.100 | 3.459 | 6 | 916336-48-6 | Yes |
| 56 | 4121732 | | 221.150 | 3.178 | 6 | 4426-24-8 | No |
| 57 | 11327649 | | 422.000 | 2.897 | 6 | 148892-98-2 | No |
| 58 | 4072412 | | 302.100 | 3.264 | 6 | 158752-98-8 | No |
| 59 | 144241 | | 262.990 | 3.962 | 6 | 67395-53-3 | No |
| 60 | 22722373 | | 332.100 | 2.614 | 6 | 154735-10-1 | No |
| 61 | 12159360 | | 837.900 | 3.357 | 6 | 223769-13-9 | No |
| 62 | 3599767 | | 307.910 | 3.422 | 6 | 659-18-7 | Yes |
| 63 | 18329245 | | 1100.100 | 3.875 | 6 | 213974-85-7 | No |
| 64 | 2736203 | | 511.900 | 3.443 | 6 | 6919-80-8 | Yes |
| 65 | 54472851 | | 607.950 | 3.432 | 6 | 755-53-3 | Yes |
| 66 | 15887311 | | 560.000 | 2.965 | 6 | 146355-12-6 | Yes |
| 67 | 52915087 | | 403.960 | 3.776 | 6 | 2317-76-2 | Yes |
| 68 | 4438527 | | 679.000 | 2.839 | 6 | 47855-94-7 | Yes |

**Table S9 | 4 final molecules**

| Compound ID | Molecular name | CAS | Formula | Price | Action |
|---|---|---|---|---|---|
| 25112575 | 2-[9,10-Di(naphthalen-2-yl)anthracen-2-yl]-4,4,5,5-tetramethyl-1,3,2-dioxaborolane | 624744-67-8 | | 240.9 ￥/g | Under preparation |
| 58050304 | 4,4,5,5-Tetramethyl-2-[10-(1-naphthyl)anthracen-9-yl]-1,3,2-dioxaborolane | 1149804-35-2 | | 557.9 ￥/g | Select to do the validation experiment |
| 58489416 | 4,4,5,5-Tetramethyl-2-[3-(triphenylen-2-yl)phenyl]-1,3,2-dioxaborolane | 1115639-92-3 | | 122.9 ￥/g | Under preparation |
| 58769449 | 4,4,5,5-Tetramethyl-2-(triphenylen-2-yl)-1,3,2-dioxaborolane | 890042-13-4 | | 68.9 ￥/g | Under preparation |

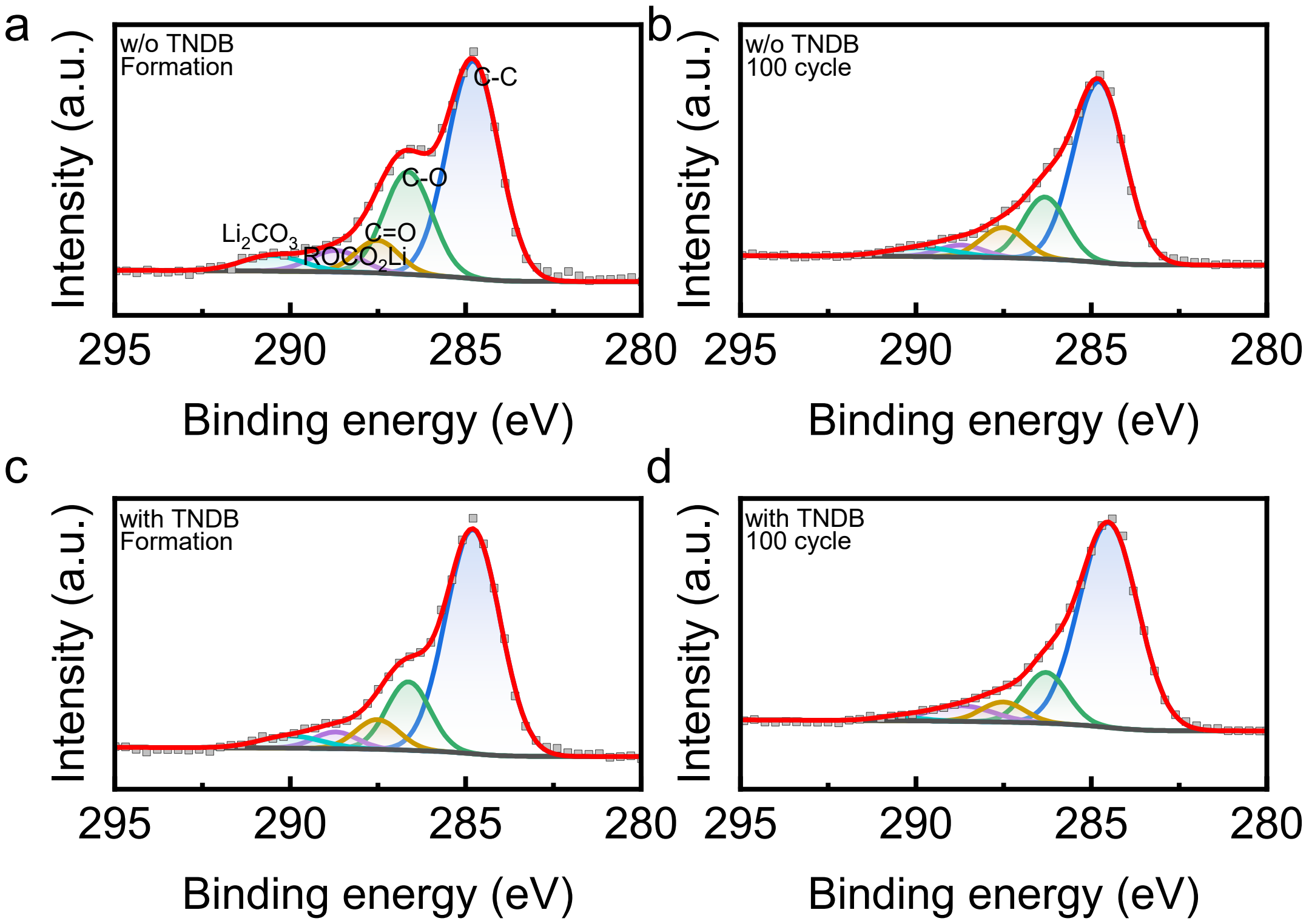

**Figure S11 | C 1s XPS analysis of graphite electrodes cycled with and without TNDB.** High-resolution C 1s spectra and peak deconvolution of graphite electrodes recovered from the baseline electrolyte after (a) formation and (b) 100 cycles, and from the TNDB-containing electrolyte after (c) formation and (d) 100 cycles at 55 °C. The fitted components are assigned to C–C, C–O, C=O, lithium alkyl carbonates ($ROCO_2Li$) and $Li_2CO_3$. Open squares represent the experimental data, red curves the overall fits and colored curves the individual fitted components. Binding energies were calibrated to the C–C peak at 284.8 eV.

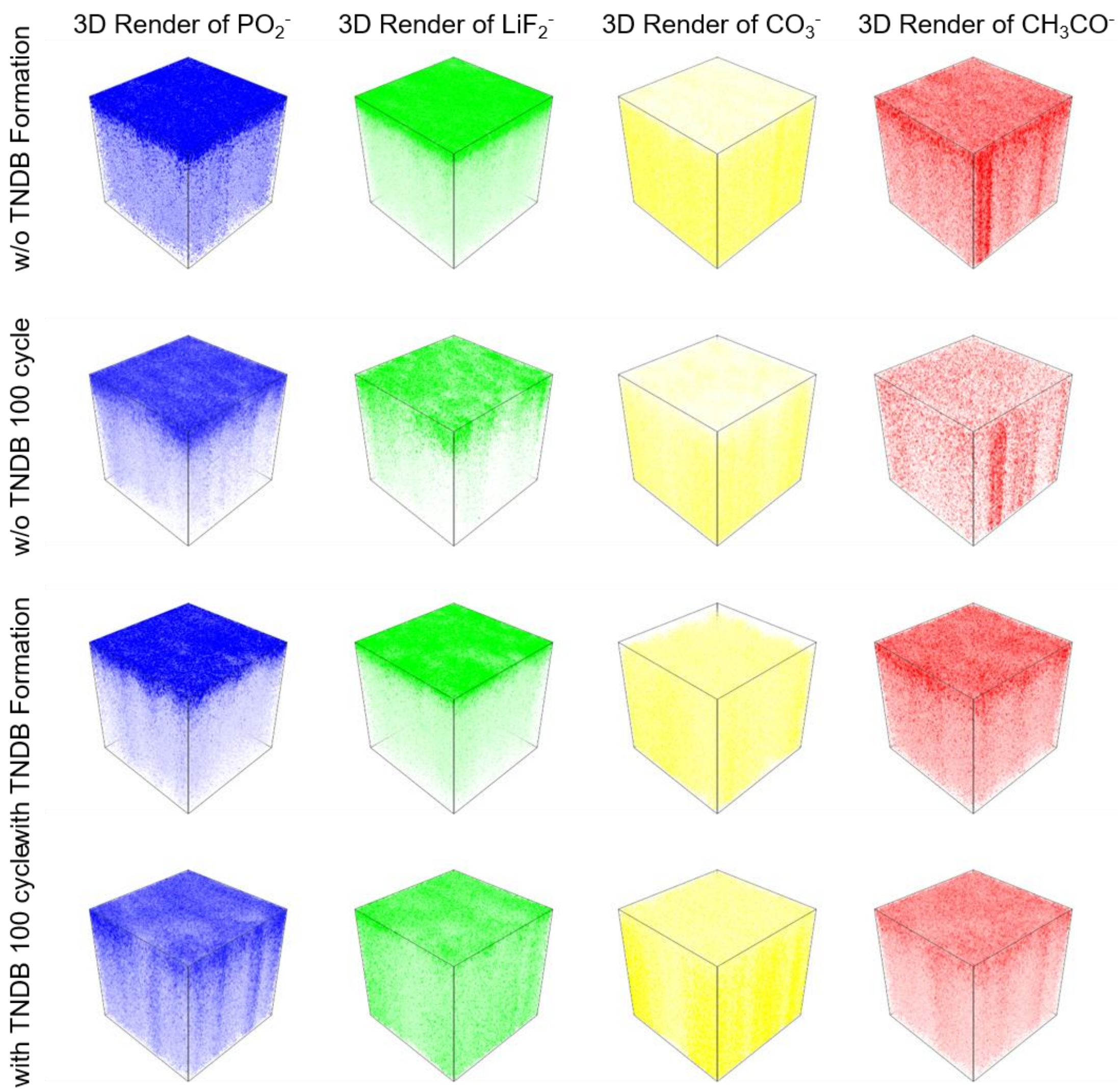

**Figure S12 | Three-dimensional ToF-SIMS compositional renderings of representative SEI fragments on graphite electrodes cycled at 55 °C with and without TNDB.** From left to right, the mapped negative secondary-ion fragments are $PO_2^-$, $LiF_2^-$, $CO_3^-$ and $CH_3CO^-$. The rows correspond, from top to bottom, to graphite electrodes recovered from the baseline electrolyte after formation and after 100 cycles, and from the TNDB-containing electrolyte after formation and after 100 cycles at 55 °C. The renderings visualize the spatial and depth-dependent distributions of the selected fragments within the sputtered interphase volume.

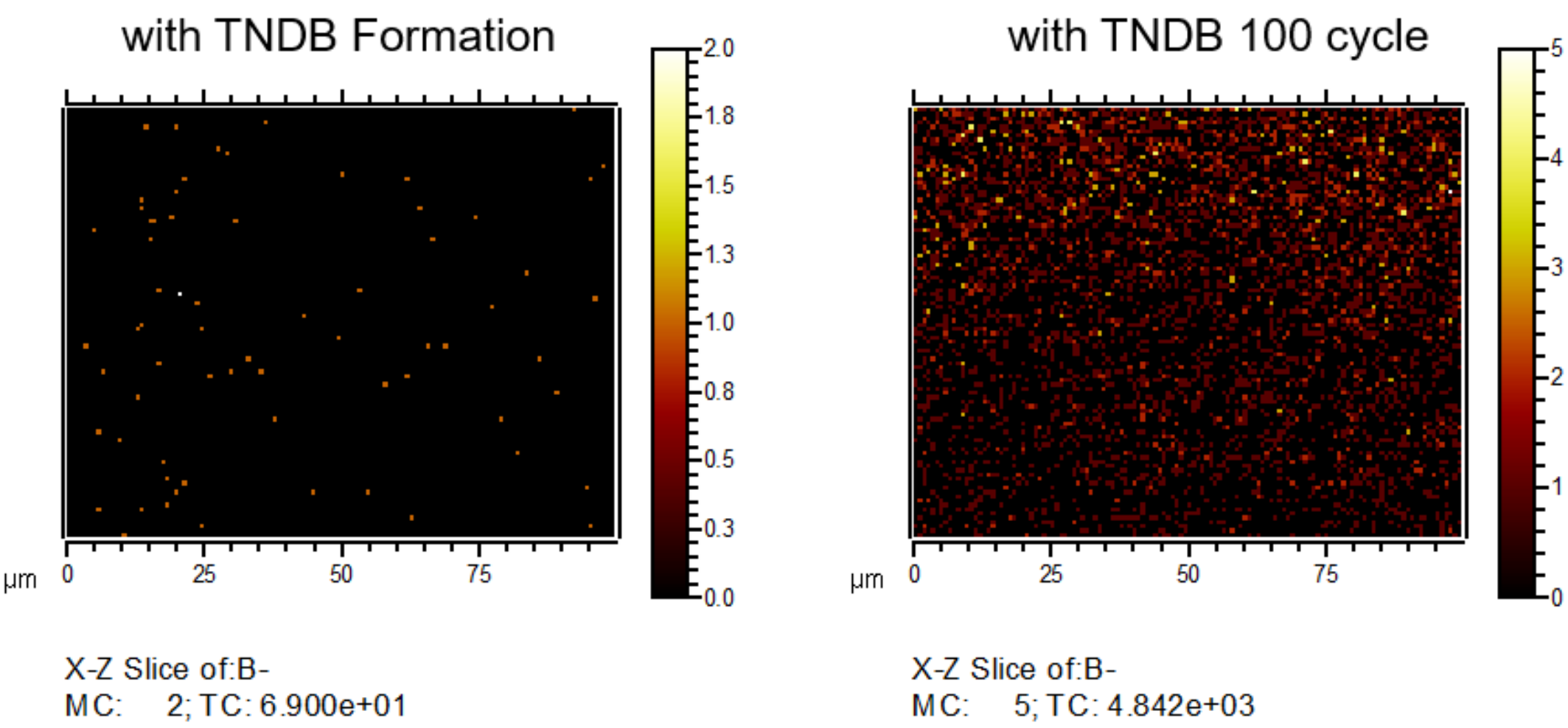

**Figure S13 | Depth-resolved distribution of additive-derived boron in the graphite SEI.** X–Z cross-sectional ToF-SIMS maps of the $B^-$ secondary-ion fragment obtained from graphite electrodes cycled in the TNDB-containing electrolyte after formation and after 100 cycles at 55 °C. Color scales are independently adjusted for visualization.

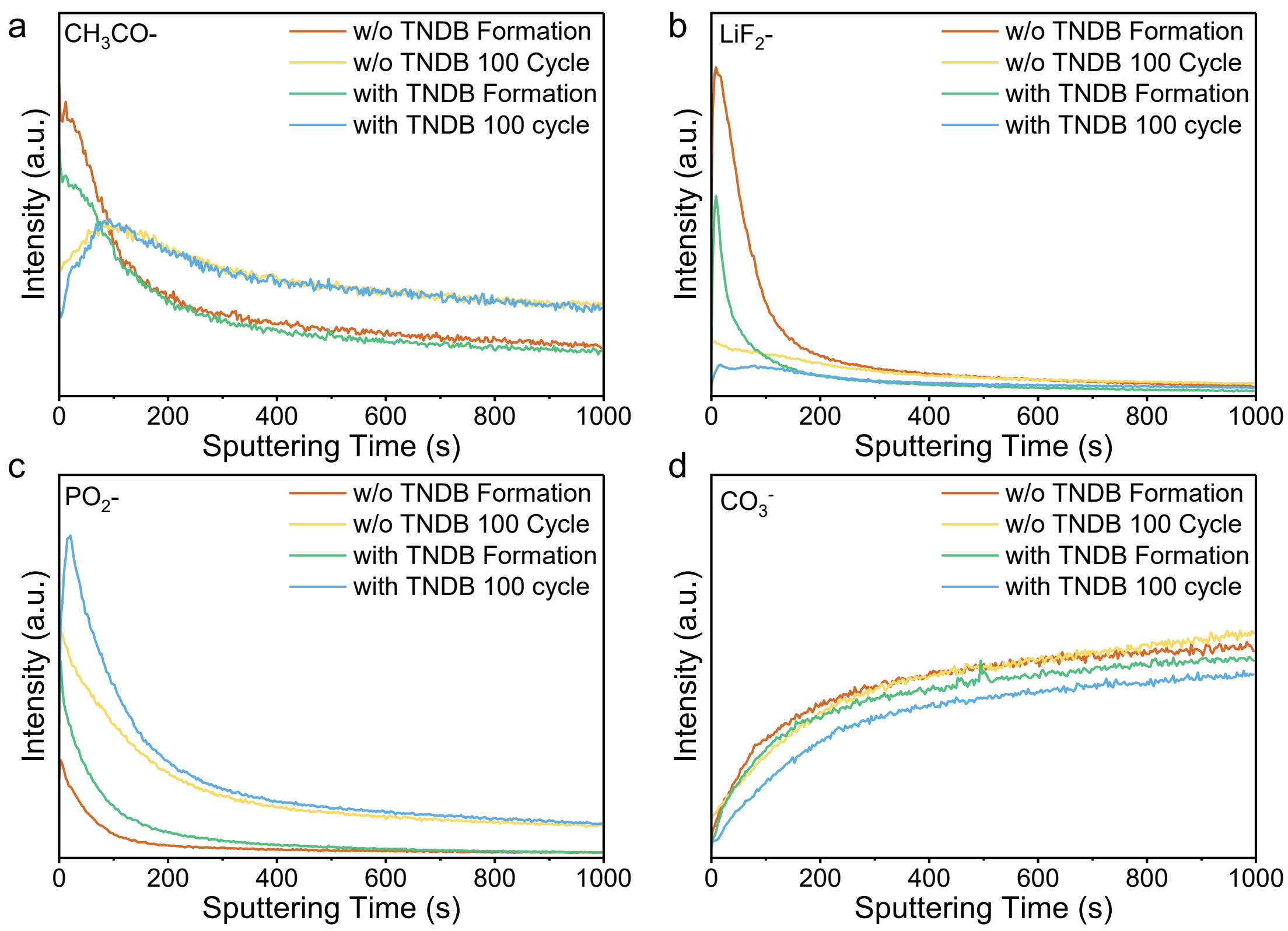

**Figure S14 | ToF-SIMS sputter-depth profiles of the graphite SEI formed with and without TNDB.** (a) $CH_3CO^-$, representing oxygenated organic fragments derived from electrolyte decomposition; (b) $LiF_2^-$, representing LiF-related inorganic species; and (c) $PO_2^-$, associated with $LiPF_6$-derived P-O species. (d) $CO_3^-$, representing carbonate-containing residues. The profiles are interpreted as relative depth-distribution trends.

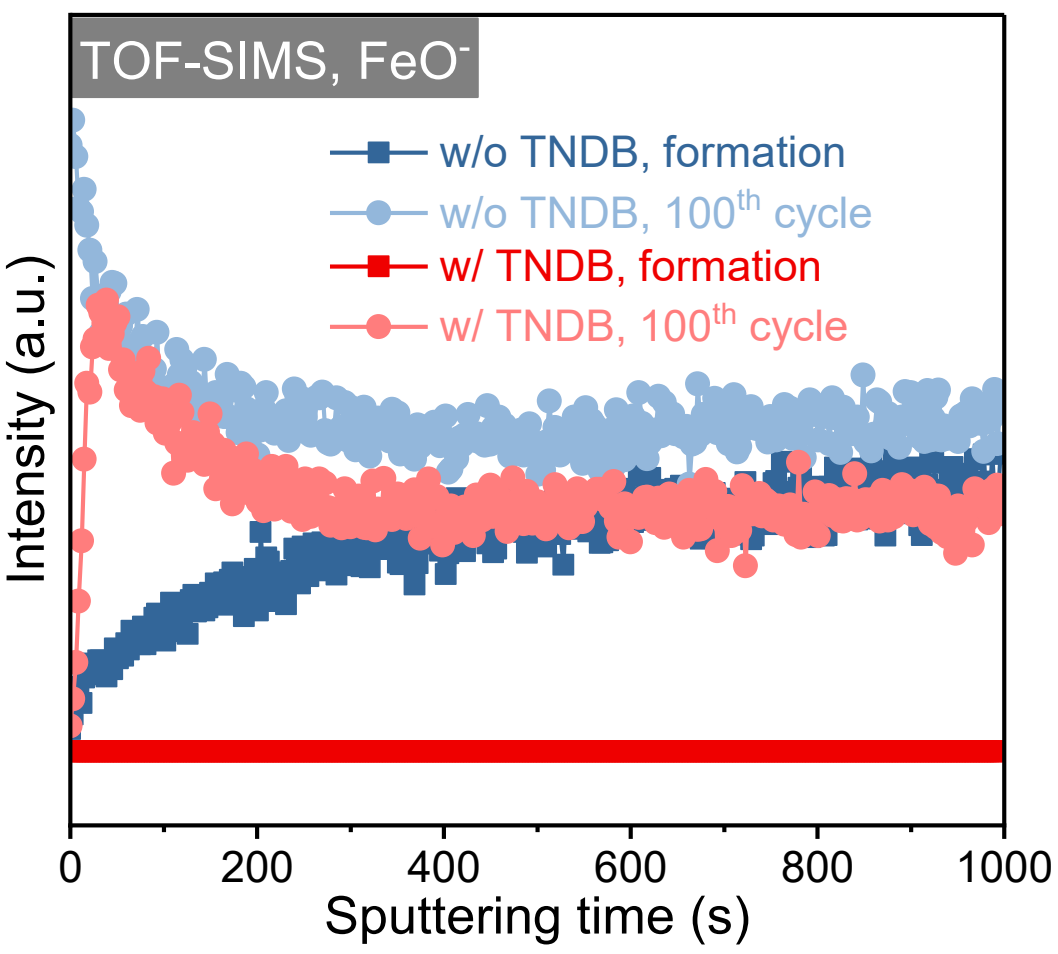

**Figure S15 | ToF-SIMS sputtering-time profiles of the $FeO^{-}$ secondary-ion fragment for graphite electrodes recovered after formation and 100 cycles.**

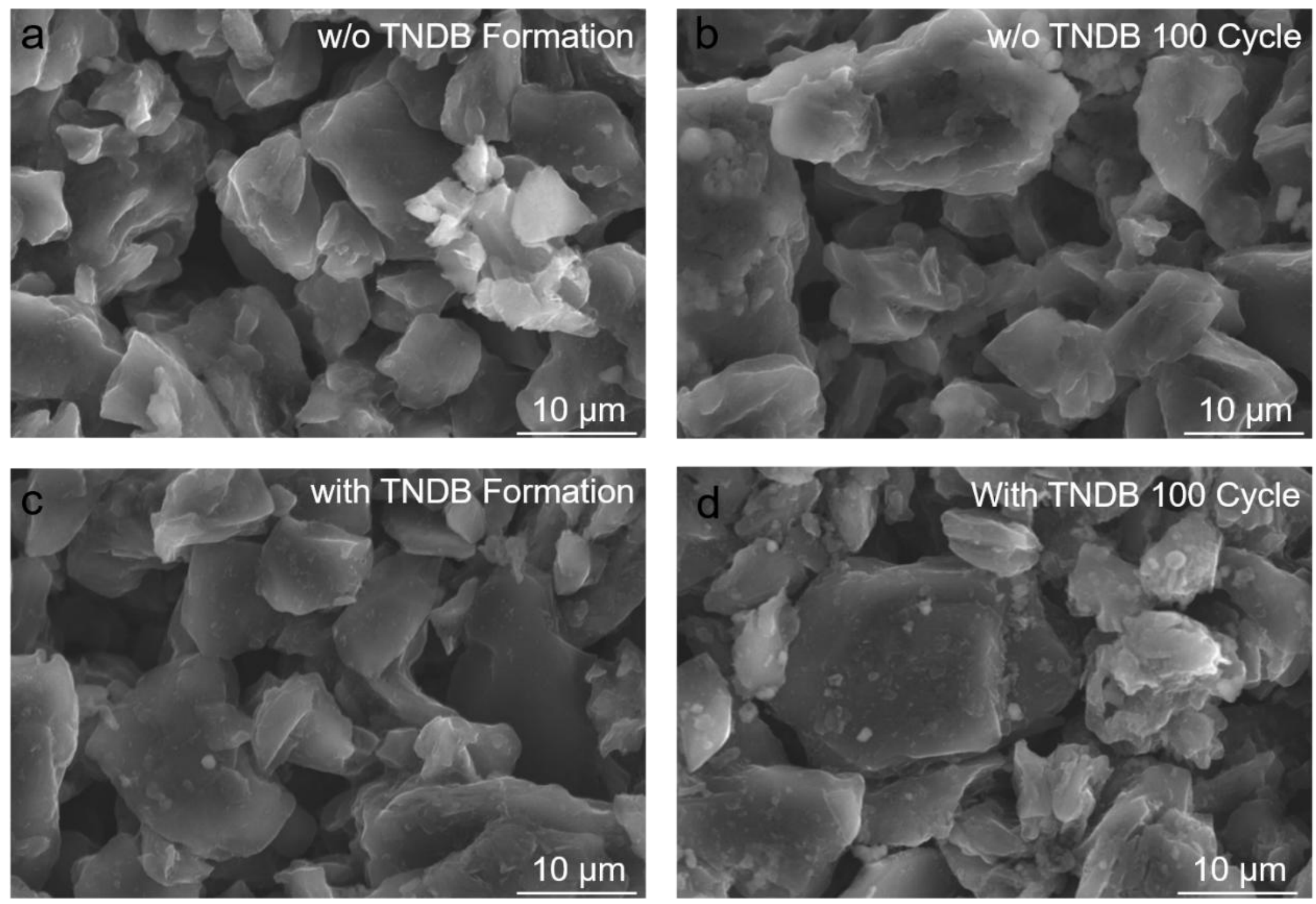

**Figure S16 | Surface morphology of graphite electrodes after cycling.** Representative SEM images of graphite electrodes recovered from $LiFePO_4$||graphite cells using the (a,b) baseline electrolyte and (c,d) TNDB-containing electrolyte after (a,c) formation and (b,d) 100 cycles at 55 °C.

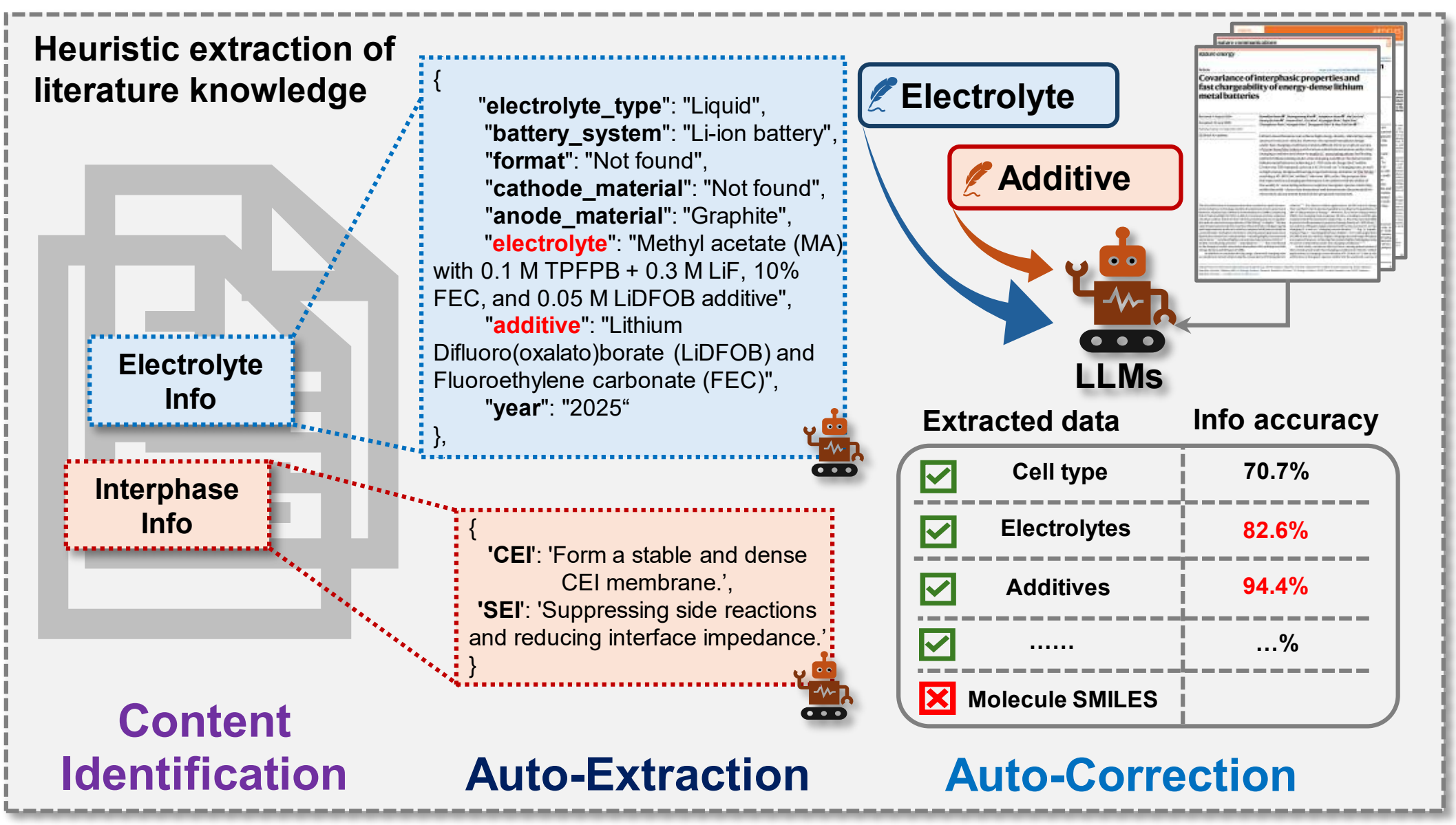

**Figure S17 | Details of the large-language-model-assisted data-mining pipeline.** Content identification: identify the electrolyte and interphase-related information. Auto-extraction: using the large language model to extract specific content guided by the user prompts. Auto-correction: read the corresponding literature for each electrolyte and additive entity again to check the correction of the extracted information. The deadline for collecting the dataset is before July 2025. And the prompts used in the extraction stage are shown in the Supplementary Figure S18 and Figure S19.

```
system_prompt = """
    The user will provide some exam text. Please parse the 'question' and 'answer' and output them in
JSON format.

    EXAMPLE INPUT:
    Suppose you are an expert in the field of batteries, here is the paper called title1, please help me
to find the following information:
    [text...]

    Please think step by step and answer the following questions.
    Question:
    1. Is the additive(s) proposed by this paper added to the battery electrolyte? Yes.
    2. What is the type of the battery electrolyte proposed by this paper? Liquid.
    3. What is the battery system tested by this paper? Li-ion battery.
    3. What is the format of the battery tested by this paper? Coin.
    4. What is the cathode material used in the battery tested by this paper? NCM811.
    5. What is the anode material used in the battery tested by this paper? Graphite.
    6. What is the electrolyte composition used in the battery tested by this paper? 1M LiPF6 in EC/EMC
(3:7 by wt.) with 2 percant FEC.
    7. What is the additive proposed by this paper, find it in the paper, including full term and
abbreviation of additives? Lithium Difluoro(oxalato)borate (LiDFOB).
    8. What is the year of publication of this paper? 2020.

    [Note]

    If the information is not found in the text, please answer 'Not found'.

    EXAMPLE JSON OUTPUT:
    {
        'title': 'title1',
        'relevant': 'Yes',
        'electrolyte_type': 'Liquid',
        'battery_system': 'Li-ion battery',
        'format': "Coin',
        'cathode_material': 'NCM811',
        'anode_material": 'Graphite',
        'electrolyte': '1M LiPF6 in EC/EMC (3:7 by wt.) with 2 percant FEC',
        'additive': 'Lithium Difluoro(oxalato)borate (LiDFOB)',
        'year': '2020'
    }
    """
```

**Figure S18 | Prompt for content identification and auto-extraction.**

```
system_prompt = """
    The user will provide some information. Please parse the 'question' and 'answer' and output them in
JSON format.

    EXAMPLE INPUT:
    Suppose you are an expert in the field of batteries, here is the electrolyte and additive
information, please help me to find the following information:
    electrolyte: [...], additive: [...]

    Please think step by step and answer the following questions.
    Quenstions:
    1. Whether the additives used in the electrolyte are appropriate for the battery's electrolyte
requirements? Yes.
    2. If this electrolyte additive is not a reasonable one, please tell me why. [Reason...]
    3. Based on your inspection, the confirmed composition of the electrolyte is? 1M LiPF6 in EC/EMC (3:7
by wt.) with 2 percant FEC.
    4. Based on your inspection, the confirmed composition of the battery electrolyte additive is?
Lithium Difluoro(oxalato)borate, Fluoroethylene carbonate.
    5. The abbreviation for the confirmed components of the battery electrolyte additive is? LiDFOB, FEC.
    6. Among the confirmed additives, those containing boron compounds are (in full name)? Lithium
Difluoro(oxalato)borate.
    7. Among the confirmed additives, those containing boron compounds are (in abbreviation)? LiDFOB.


    [Note]
    If the electrolyte additive is a reasonable one, please answer 'Reasonable' to question 2.
    If question 1 is 'No', please answer question 2, and skip questions 3, 4, 5, 6, and 7, answering them
as 'Not found'.


    EXAMPLE JSON OUTPUT:
    {
        'appropriate': 'Yes',
        'reason': 'Reasonable',
        'electrolyte': '1M LiPF6 in EC/EMC (3:7 by wt.) with 2 percant FEC',
        'additive_full_name': 'Lithium Difluoro(oxalato)borate, Fluoroethylene carbonate',
        'additive_abbr_name': 'LiDFOB, FEC',
        'boron_additive_full_name': 'Lithium Difluoro(oxalato)borate'
        'boron_additive_abbr_name': 'LiDFOB'
    }
    """
```

**Figure S19 | Prompt for auto-correction.**

**Table S10 | Edge features and corresponding physical meanings**

| Feature index | Physical Meaning | Feature type | Encoding type | Data type |
|---|---|---|---|---|
| 1 | bond_type_single | Edge feature | One hot encoding | String |
| 2 | bond_type_double | Edge feature | One hot encoding | String |
| 3 | bond_type_triple | Edge feature | One hot encoding | String |
| 4 | bond_type_aromatic | Edge feature | One hot encoding | String |
| 5 | bond_is_conjugated | Edge feature | Numerical | Int |
| 6 | bond_is_in_ring | Edge feature | Numerical | Int |
| 7 | bond_dir_none | Edge feature | One hot encoding | String |
| 8 | bond_dir_beginwedge | Edge feature | One hot encoding | String |
| 9 | bond_dir_begindash | Edge feature | One hot encoding | String |
| 10 | bond_dir_enddownright | Edge feature | One hot encoding | String |
| 11 | bond_dir_endupright | Edge feature | One hot encoding | String |
| 12 | bond_is_aromatic | Edge feature | Numerical | Int |
| 13 | stereo_type_stereoz | Edge feature | One hot encoding | String |
| 14 | stereo_type_stereoe | Edge feature | One hot encoding | String |
| 15 | stereo_type_stereoany | Edge feature | One hot encoding | String |
| 16 | stereo_type_stereonone | Edge feature | One hot encoding | String |